\documentclass[11pt]{article}
\usepackage{mathrsfs}
\usepackage{latexsym}
\usepackage{amsmath,amssymb,amsfonts, amsthm}
\usepackage[pdftex]{hyperref}
\usepackage{amsthm}
\usepackage{amsfonts,cite}
\usepackage[usenames]{color}
\usepackage{amssymb}
\usepackage{graphicx}
\usepackage{amsmath}
\usepackage{amsfonts}
\usepackage{amsthm}
\usepackage{mathrsfs}
\usepackage{dsfont}
\usepackage{hyperref}
\usepackage{pgfplots}
\usepackage{geometry}
\usepackage{booktabs}
\usepackage{hyperref}
\newcommand\be{\begin{equation}}
\newcommand\ee{\end{equation}}
\newcommand\ber{\begin{eqnarray}}
\newcommand\eer{\end{eqnarray}}
\newcommand\berr{\begin{eqnarray*}}
\newcommand\eerr{\end{eqnarray*}}
\newcommand\bea{\begin{eqnarray}}
\newcommand\eea{\end{eqnarray}}
\newcommand\ba{\begin{array}}
\newcommand\ea{\end{array}}
\newcommand\bfR{\mathbb{R}}
\newcommand\dd{\mathrm{d}}

\newcommand\ii{\mbox{i}}
\newcommand\e{\mathrm{e}}
\newcommand\eq{\eqref}\newcommand\lb{\label}

\newcommand\ri{\mathrm{i}}

\newcommand\pa{\partial}
\newcommand\kb{k_{\rm{B}}}

\newcommand{\nn}{\nonumber}

\newcommand{\keywords}[1]{\par\noindent\textbf{Keywords:} #1\par}
\newcommand\Om{\Omega}

\newcommand{\bi}{\begin{itemize} }
  \newcommand{\ei}{\end{itemize} }

\newtheoremstyle{mythm}{1.5ex plus 1ex minus .2ex}{1.5ex plus 1ex
minus .2ex}{\kai}{\parindent}{\song\bfseries}{}{1em}{}
\numberwithin{equation}{section}\numberwithin{figure}{section}

\allowdisplaybreaks[4]
\begin{document}
\title{Bogomol'nyi Equations in Two-Species Born--Infeld \\Theories  Governing Vortices and Antivortices}
\author{Aonan Xu \\ School of Mathematics and Statistics\\Henan University\\
Kaifeng, Henan 475004, P. R.  China\\ \\ Yisong Yang \\ Courant Institute of Mathematical Sciences\\ New York University\\New York,  New York 10012, USA}
\date{}
\maketitle

\begin{abstract}
We derive several new Bogomol'nyi (self-dual) equations in two-species
$U(1)\times U(1)$ gauge theories governed by the Born--Infeld nonlinear electrodynamics.
By identifying appropriate Born--Infeld type Higgs potentials, we show that the highly
nonlinear energy functionals admit exact topological lower bounds saturated by coupled
first-order equations. The resulting models accommodate both vortex-vortex and
vortex-antivortex configurations and generalize previously known single-species
Born--Infeld systems to interacting multi-component settings.

Beyond the derivation of the Bogomol'nyi equations, we develop an exact thermodynamic
theory for pinned multivortex configurations in both the full plane and compact doubly
periodic domains. Owing to the linear dependence of the Bogomol'nyi energy spectrum on
topological charges, we obtain closed-form expressions for the canonical partition function,
internal energy, heat capacity, and magnetization. In compact domains, the Bradlow type
geometric bounds constrain admissible vortex numbers and lead to qualitatively new
high-temperature behavior. In particular, vortex-only systems exhibit spontaneous
magnetization, while vortex-antivortex systems do not, reflecting the underlying symmetry
between opposite topological charges. These results provide a rare analytically solvable
framework for studying thermodynamics in nonlinear multi-component gauge theories
regulated by the Born--Infeld electrodynamics.

%Nonlinear structures in the Born--Infeld type theories make the construction of solutions to the governing equations a daunting task. The availability of the Bogomol'nyi reductions would therefore shed considerable light on the understanding of such theories. The main contribution of this paper is the derivation of several new Bogomol'nyi equations in the Born--Infeld models within a product Abelian gauge field theory framework. The models investigated include both vortex systems and vortex-antivortex systems. Moreover, we present a complete thermodynamic analysis and obtain exact thermodynamic results for such multivortex systems in both the full-plane setting and on a compact periodic lattice domain. These include closed-form canonical partition functions, internal energy, heat capacity, and magnetization, revealing distinct high- and low-temperature behaviors, the Meissner effect at zero temperature, and the influence of the Bradlow type geometric bounds on vortex condensation. The discovery of these Bogomol'nyi equations in multi-component Born--Infeld theories provides an analytical window into nonlinear gauge systems of high physical and mathematical complexity.

\medskip

\keywords{Born--Infeld theory,
          gauge field theories,
          vortices,
          topological defects,
          Bogomol'nyi equations,
          superconductivity,
          thermodynamics,
          magnetic properties}

\medskip

\noindent{\bf PACS numbers}. 02.30.Jr, 02.30.Xx, 11.15.-q, 74.25.Ha

\medskip

\noindent{\bf MSC numbers}. 35J50, 53C43, 58E15, 81T13, 82B26

\end{abstract}
\section{Introduction}
\setcounter{equation}{0}

The governing equations of gauge field theories in all dimensions are known to be difficult to solve exactly due to their complicated nonlinear structures. In 1976, Bogomol'nyi \cite{Bo} discovered a variational structure in the energy functionals in the two- and three-dimensional Abelian Higgs theory and the $SU(2)$ Yang--Mills theory, respectively, which renders a first-order equation
reduction from the original second-order equations, such that the solutions saturate the topological lower bounds of the energies and give rise to multiply centered solitons of the nature of
field-theoretical particles \cite{JT}. As a by-product, the work of Bogomol'nyi also yields a variational interpretation of the earlier explicit construction of
Prasad and Sommerfield \cite{PS} of the monopole and dyon solutions of the $SU(2)$-theory in the zero-coupling constant limit and shares the self-dual structure of the Yang--Mills instantons
in four dimensions conceptualized by Belavin, Polyakov, Schwartz, and Tyupkin \cite{BPST} in the same variational-structure sense that their self-dual equations may be viewed as
a Bogomol'nyi equation reduction of the original second-order Yang--Mills equations. For this reason, the Bogomol'nyi equations are generally referred to as the self-dual equations. In fact, Witten
later showed that \cite{Witten}, with an appropriate ansatz, the self-dual Yang--Mills equations may be realized by a system of Bogomol'nyi's equations in a two-dimensional Abelian Higgs model in the same
spirit as in \cite{Bo} but now over the Poincar\'{e} half-plane of the hyperbolic space characteristic and the instantons are represented by vortices. In contemporary theoretical physics research,
the exploration of the Bogomol'nyi structures has led to the in-depth understanding of numerous otherwise rather intractable field equations arising in the electroweak theory \cite{AO1,AO2,AO3,AO4,SYew1,SYew2,BL1,BL2,Yew}, Chern--Simons models \cite{HKP,JW,Caff-Y,Han-L-Y,Han-Y2}, cosmic strings \cite{V,CG,Ycosmic1,Ycosmic2,Ybook}, supersymmetric gauge theories in connection with the monopole confinement
mechanism through the colored Meissner effect and vortex condensation \cite{Auzzi,Konishi1,Konishi2,Chen-Yang,Lieb-Yang,Han-Yang}, and the vortices  arising in the Born--Infeld electrodynamics
\cite{BI1,BI2,SH,Ybi,Han}. In these models, since the minimal energies are explicitly stratified in terms of the vortex numbers given as topological charges, the associated  partition functions can be evaluated
exactly so that the vortex thermodynamic properties become describable.

In condensed-matter physics, a product Abelian Higgs theory with the natural $U(1)\times U(1)$ gauge symmetry naturally describes systems possessing two independent complex condensate fields or order parameters each associated with a distinct electronic band or condensate species.
Such two-gap or multi-band superconductors or superfluids (e.g. MgB$_2$, iron-based superconductors, liquid metallic hydrogen, etc) are characterized by interband couplings and partial magnetic flux quantization.
In particle and cosmological models, product Abelian gauge groups arise in electroweak extensions, hidden-photon or dark-sector couplings, and cosmic-string constructions.
In such settings, each condensate couples to its own gauge potential or to distinct components of the electromagnetic fields, leading to the magnetic flux quantization of its own signature and contributing to vortex molecules---composite vortices with sub-quantum flux and nontrivial interaction potentials. See \cite{AV,Bab1,Bab2,Bab3,Nitta,HK} and literature therein for development.
Establishing Bogomol'nyi type equations in such multi-component settings not only yields exact analytical insight but also clarifies the interplay between distinct flux sectors and the formation of mixed vortex--vortex and vortex--antivortex structures. In \cite{Schroers}, a systematic derivation of the Bogomol'nyi equations is carried out by
Schroers for the general $m$-fold product of $U(1)$, or $U(1)^m$, Abelian Higgs theory, and an existence and uniqueness theorem for the solutions of the equations is established in \cite{Yjfa}. In \cite{TW}, Tong and Wong
propose a $U(1)\times U(1)$ gauge field theory and link it to the study of magnetic impurities. Their model shares the same gauge group structure considered here, but focuses on impurity interactions rather than the Born--Infeld nonlinear electrodynamics and thermodynamic properties that we investigate. See also \cite{HY0}.
In \cite{HY}, such a two-species theory is extended to include the interaction of vortices
and antivortices along the line of the studies \cite{Schroers2,Yprl,Yvav,SSY}.

When the electromagnetic sector is replaced by the Born--Infeld type nonlinear electrodynamics, the interplay between the gauge field and the Higgs scalar field becomes even more intricate.
The nonlinear field strength saturation introduces a natural bound on the associated magnetic field and leads to finite-energy vortex structures introduced by the
Bogomol'nyi type first-order equations when the Higgs field potential assumes a Born--Infeld form as discovered by Shiraishi and Hirenzaki \cite{SH}.
Because of its radical-root form, obtaining exact Bogomol'nyi reductions within Born--Infeld type systems is far from trivial.
When such reductions do occur, they identify distinguished sectors where the nonlinearity remains compatible with self-duality and energy minimization.
The resulting equations thus carry both analytical elegance and physical significance, connecting field regularization with topological quantization \cite{Han,LY}. In particular, in
\cite{LY}, the Bogomol'nyi equations are derived in the Born--Infeld electrodynamics framework coupled with a harmonic map scalar field accommodating coexisting vortices and
antivortices, extending the line of results obtained in \cite{Schroers,Yprl,Yvav,SSY} subject to the Maxwell type electrodynamics.

As described, studying vortices in two-species $U(1)\times U(1)$ Abelian Higgs models illuminates how multiple gauge and scalar degrees of freedom conspire to produce composite, fractional, and mixed-flux topological excitations. These models bridge condensed-matter, cosmological, and string-theoretic physics, serving as versatile laboratories for exploring the interplay between nonlinearity, topology, and gauge symmetry.
When combined with the Born--Infeld type electrodynamics, they further reveal how nonlinear field regularization modifies vortex structure and stability, making the appearance of the Bogomol'nyi equations both surprising and profound. With this motivation,
in this paper, we develop and analyze a class of $U(1)\times U(1)$ Born--Infeld--Higgs models admitting new Bogomol'nyi equations that couple two gauge fields and two charged scalar fields.
The models generalize previously known single-species Born--Infeld vortex and vortex-antivortex systems to a two-species framework, allowing interactions between distinct magnetic sectors.
Through appropriate choices of potentials, we demonstrate that the energy functionals admit  lower bounds saturated by self-dual configurations satisfying some coupled first-order equations.
These equations govern either vortex-vortex and vortex-antivortex solutions, depending on the models. For the vortex and vortex-antivortex systems derived, we shall obtain explicit quantized formulas for the magnetic fluxes and total energies. In the compact domain settings, we also systematically derive various Bradlow bounds which constrain vortex numbers or topological charges
by the geometry of the systems.
Furthermore, beyond the derivation of the Bogomol'nyi equations themselves, we investigate the thermodynamic properties of the resulting vortex and vortex-antivortex systems under the simplifying assumption of pinned vortex cores---a setting relevant to type-II superconductors with strong pinning, optical lattices, or engineered vortex arrays. For two representative models---a two-species vortex-only system and a two-species vortex-antivortex system---we evaluate the canonical partition functions explicitly in both the full plane and on a compact doubly periodic domain. The Bogomol'nyi structure guarantees that the energy spectrum is linear in the topological quantum numbers, allowing exact calculation of thermodynamic quantities. Our analysis yields closed-form expressions for the internal energy, heat capacity, and magnetization as functions of temperature and applied magnetic field. Notably, we recover the Meissner effect at zero temperature in all models, and we uncover qualitatively distinct high-temperature behaviors: in the full plane the heat capacity approaches a nonzero constant, whereas on a compact domain it may vanish or saturate at a nonzero level, characterized by whether Bradlow type geometric bounds are enforced. Moreover, spontaneous magnetization occurs in vortex-only systems but is absent in vortex-antivortex systems---a direct consequence of the symmetry between vortices and antivortices in the energy spectrum. These results provide a solvable statistical-mechanical framework for studying thermodynamic phases of multicomponent topological matter and reveal how geometry and topology jointly shape the response of nonlinear gauge systems.
The discovery of these Bogomol'nyi equations in multi-component Born--Infeld theories, including the distinctive thermodynamic properties of the vortex systems in various settings,  provides an analytical window into nonlinear gauge systems of high physical and mathematical complexity characterized by the Born--Infeld electrodynamics.
To our knowledge, such an explicit and exact thermodynamic treatment of pinned Bogomol'nyi vortices in multi-component Born--Infeld theories has not been systematically presented in the literature.

The principal results of this work may be summarized as follows:
\begin{enumerate}
\item We derive new Bogomol'nyi equations in two-species $U(1)\times U(1)$ gauge theories
with the Born--Infeld electrodynamics, covering both vortex-only and vortex-antivortex systems.
\item We obtain exact quantization formulas for magnetic fluxes and minimum energies in all
models, expressed solely in terms of topological charges.
\item On compact periodic domains, we systematically establish the Bradlow type bounds that
constrain admissible vortex numbers and depend explicitly on the geometry of the domain.
These bounds depend explicitly on the Born parameters,
which act as nonlinear regulators controlling vortex accommodation.
\item For pinned vortex configurations, we compute exact canonical partition functions and
derive closed-form expressions for internal energy, heat capacity, and magnetization.
\item We identify spontaneous magnetization in vortex-only systems and demonstrate its
absence in vortex-antivortex systems as a consequence of topological charge symmetry.
\end{enumerate}

Here is an outline of this work. In the next section, we recall two single-species Born--Infeld models which inspire our present work and also place our study in perspectives. In Sections \ref{sec3} and \ref{sec4},  we develop two-species Born--Infeld and Maxwell--Born--Infeld vortex models, respectively. In Sections \ref{sec5} and \ref{sec6}, we study two-species Born--Infeld and Maxwell--Born--Infeld models which accommodate vortex and antivortex pairs, respectively. In Sections \ref{sec7} and \ref{sec8}, we consider two mixed-type models: the Maxwell vortices and Born--Infeld vortices-antivortices, the Born--Infeld vortices and Maxwell vortices-antivortices, respectively. In Sections \ref{sec9} and \ref{sec10}, we present a series of thermodynamic results for two typical models in the full plane and  bounded periodic lattice domain settings, respectively. In particular, in the latter setting, we present a series of
novel thermodynamic features associated with the Bradlow bounds.
In Section \ref{sec11}, we establish the Bradlow bounds for the other four models which lay a foundation for similar thermodynamic explorations of the systems of pinned vortices and antivortices coupled together.
In Section \ref{sec12}, we draw conclusions and make comments.

\section{Bogomol'nyi equations in single-species models}\lb{sec2}

This section serves as a preparatory review and motivation for the developments that follow. We recall two representative single-species models in which the Born--Infeld nonlinear electrodynamics admits a Bogomol'nyi reduction: the Born--Infeld Abelian Higgs vortex model and the gauged harmonic map model supporting vortices and antivortices. These examples illustrate how a carefully chosen potential restores a self-dual structure despite the radical nonlinearity of the Born--Infeld action. By presenting these models in a unified and self-contained manner, we fix notation, highlight the role of the Born--Infeld nonlinearity, and clarify the mechanism by which topological energy bounds and first-order equations emerge. This discussion provides a conceptual and technical foundation for the multi-species generalizations developed in subsequent sections.

Specifically, our work arises from the Born--Infeld Abelian Higgs model of Shiraishi and Hirenzaki \cite{SH} for vortices and the harmonic map model \cite{LY} accommodating both vortices and antivortices.

Coupled with a complex scalar Higgs field $\phi$,
the simplest model of our interest governed by the Born--Infeld electrodynamics is defined by the Lagrangian action density \cite{BI2,SH,Ybi,Y-AOP}
\bea
&&{\cal L}=b^2\left(1-\sqrt{1-\frac2{b^2} s}\right)+\frac12 D_{\mu}\phi\overline{D^{\mu}\phi}-V(|\phi|^2),\lb{2.1}\\
&&s=-\frac14F_{\mu\nu}F^{\mu\nu}+\frac{\kappa^2}{32}\left(F_{\mu\nu}\tilde{F}^{\mu\nu}\right)^2,\lb{2.2}
\eea
over the Minkowski spacetime $\bfR^{3,1}$ equipped with the metric $\eta^{\mu\nu}=\mbox{diag}\{1,-1,-1,-1\}$, which is used to lower or raise indices,
where the electromagnetic tensor field $F_{\mu\nu}=\pa_\mu A_\nu-\pa_\nu A_\mu$ and its Hodge dual $\tilde{F}^{\mu\nu}=\frac12\epsilon^{\mu\nu\alpha\gamma}F_{\alpha\gamma}$ are
associated with the underlying electric field ${\bf E}=(E^i)$ and magnetic field ${\bf B}=(B^i)$ through the expressions
\be\lb{2.3}
(F^{\mu\nu})=\left(\begin{array}{cccc}0&-E^1&-E^2&-E^3\\E^1&0&-B^3&B^2\\E^2&B^3&0&-B^1\\E^3&-B^2&B^1&0\end{array}\right),\quad
(\tilde{F}^{\mu\nu})=\left(\begin{array}{cccc}0&-B^1&-B^2&-B^3\\B^1&0&E^3&-E^2\\B^2&-E^3&0&E^1\\B^3&E^2&-E^1&0\end{array}\right),
\ee
induced from a real-valued gauge field $A_\mu$,
$b>0$ is the usual Born parameter,  and $\kappa>0$ is an electromagnetic coupling parameter that directly mixes the interaction of the fields ${\bf E}$ and ${\bf B}$,
$D_{\mu} \phi =\pa_{\mu} \phi -\ri A_{\mu} \phi$ is the gauge-covariant derivative, and $V(|\phi|^2)$ is a potential density term that introduces a spontaneous symmetry breaking
mechanics with the normalized  property $V(\xi)=0$ (say) where $\xi>0$ reflects the scale of broken symmetry.

With \eq{2.1} and \eq{2.2}, the Euler--Lagrange equations are
\bea
D_{\mu}D^{\mu}\phi&=&-2V^{\prime}(|\phi|^2)\phi,\lb{ax2.4} \\
\pa_\mu P^{\mu\nu}&=& -\frac\ri2(\phi \overline{D^{\nu} \phi}-\overline{\phi}D^{\nu}\phi),\lb{ax2.5}
\eea
where
\be\lb{2.6}
P^{\mu\nu}=\frac1{\sqrt{1-\frac{2}{b^2} s}}\left(F^{\mu\nu}-\frac{\kappa^2}4(F_{\alpha\beta}\tilde{F}^{\alpha\beta})\tilde{F}^{\mu\nu}\right)
\ee
gives rise to the electric displacement field and magnetic intensity field, which do not concern us here.

In this work, we shall focus on magnetic vortices and antivortices such that the fields depend only on the planar coordinates $x_1, x_2$ and are independent of the time coordinate $x_0$ and the vertical
coordinate  $x_3$ of the space and that $A_0=A_3=0$. In this situation, the equations \eq{ax2.4} and \eq{ax2.5} become
\bea
D_i^2 \phi&=&2V'(|\phi|^2)\phi,\lb{ax2.7}\\
\pa_j\left(\frac{ F_{ij}}{\sqrt{1+\frac{F_{12}^2}{b^2}}}\right)&=&\frac{\ri}2(\phi\overline{D_i\phi}-\overline{\phi}D_i\phi),\lb{ax2.8}
\eea
where the summation convention is observed in repeated indices $i,j=1,2$. It is clear that the equations \eq{ax2.7} and \eq{ax2.8} are the Euler--Lagrange equations of the reduced Hamiltonian
energy density
\be
\mathcal{H}=-{\cal L}=b^2\left(\sqrt{1+\frac{F_{12}^2}{b^2}}-1\right)+\frac12 |D_i\phi|^2+V(|\phi|^2).\lb{ax2.9}
\ee
The equations \eq{ax2.7} and \eq{ax2.8} are hard to solve. Fortunately, extending the method of Bogomol'nyi \cite{Bo}, Shiraishi and Hirenzaki \cite{SH} are able to rewrite the Hamiltonian \eq{ax2.9} as
\bea\lb{ax2.10}
{\cal H}
&=&\frac1{2\sqrt{1+\frac{F_{12}^2}{b^2}}}\left(F_{12}\pm\sqrt{1+\frac{F_{12}^2}{b^2}}\,\frac12(|\phi|^2-\xi)\right)^2\mp\frac12 F_{12}(|\phi|^2-\xi)-b^2\nn\\
&&+\frac{b^2}{2\sqrt{1+\frac{F_{12}^2}{b^2}}}\left(\sqrt{1+\frac{F_{12}^2}{b^2}}\sqrt{1-\frac1{4b^2}(|\phi|^2-\xi)^2}-1\right)^2+b^2\sqrt{1-\frac1{4b^2}(|\phi|^2-\xi)^2}\nn\\
&&+\frac12|D_1\phi\pm\ii D_2\phi|^2\pm\frac\ii2 (D_1\phi \overline{D_2\phi}-\overline{D_1\phi}D_2\phi)+V(|\phi|^2),
\eea
where $\xi>0$ is a constant and we have used the identity
\be\lb{ax2.11}
|D_1\phi|^2+|D_2\phi|^2=|D_1\phi\pm\ii D_2\phi|^2\pm \ri(D_1\phi \overline{D_2\phi}-\overline{D_1\phi}D_2\phi).
\ee
Thus, with the commutator relation
\be
(D_1D_2-D_2 D_1)\phi=-\ii F_{12}\phi,
\ee
and the current, $J_i$,  defined by the right-hand side of \eq{ax2.8}, we have the associated vorticity field
\be\lb{ax2.13}
J_{12}=\pa_1 J_2-\pa_2 J_1=\ii (D_1\phi\overline{D_2\phi}-\overline{D_1\phi} D_2\phi)-|\phi|^2 F_{12}.
\ee
Consequently, if the potential energy density assumes the critical expression
\be\lb{ax2.14}
V(|\phi|^2)=b^2\left(1-\sqrt{1-\frac1{4b^2}(|\phi|^2-\xi)^2}\right),
\ee
then we can use \eq{ax2.13} and \eq{ax2.14} in \eq{ax2.10} to get
\be\lb{ax2.15}
{\cal H}\geq \pm \frac12 (\xi F_{12}+J_{12}).
\ee
It is interesting that \eq{ax2.14} indeed accommodates a spontaneous symmetry breaking structure with $V(\xi)=0$ and is asymptotically of the Bogomol'nyi form \cite{Bo,JT}
\be
V(|\phi|^2)\approx \frac18 (|\phi|^2-\xi)^2,\quad b\gg1.
\ee
Since the finite-energy condition implies that $D_i\phi$ vanishes at infinity of $\bfR^2$ exponentially fast, we have $\int_{\bfR^2} J_{12}\,\dd x=0$. On the other hand, the quantity
$F_{12}$ is topologically the first Chern density whose normalized integral is an integer,
\be
\frac1{2\pi}\int_{\bfR^2} F_{12}\,\dd x=N,
\ee
known as the vortex number, we get from \eq{ax2.15} the energy lower bound
\be
E=\int_{\bfR^2}{\cal H}\,\dd x\geq\xi \pi |N|,
\ee
which is saturated when the Bogomol'nyi type equations
\bea
F_{12}\pm\frac12 \sqrt{1+\frac{1}{b^2}F_{12}^2}(|\phi|^2-\xi)&=&0,\lb{x1.4}\\
\sqrt{1+\frac{1}{b^2}F_{12}^2} \sqrt{1-\frac{1}{4b^2}(|\phi|^2-\xi)^2}&=&1,\lb{x1.5}\\
D_1\phi\pm \ri D_2 \phi &=& 0,\lb{x1.6}
\eea
hold \cite{SH}
according to $N=\pm|N|$. It is clear that the system of equations \eq{x1.4} and \eq{x1.5} can be recast into the equivalent single equation
\be\lb{ax2.22}
F_{12}=\pm\frac{(\xi-|\phi|^2)}{2\sqrt{1-\frac{1}{4b^2}(|\phi|^2-\xi)^2}}.
\ee

Hence the Bogomol'nyi equations are reduced into the coupled system of equations \eq{x1.6} and \eq{ax2.22}. It is well known \cite{JT} that \eq{x1.6} implies that the zeros of the field $\phi$ are
isolated with integer multiplicities, say $q_1,\dots, q_n$, listed repeatedly to count for multiplicities, such that the substitution $u=\ln|\phi|^2$ renders the system
into the nonlinear elliptic equation
\be\lb{ax2.23}
\Delta u=\frac{\e^u-\xi}{\sqrt{1-\frac1{4b^2}(\e^u-\xi)^2}}+4\pi\sum_{s=1}^N \delta_{q_s}(x),\quad x\in\bfR^2,
\ee
where $\delta_q(x)$ is the Dirac measure concentrated at $q\in\bfR^2$.

Furthermore, in a similar manner, the gauged harmonic map model in which electromagnetism is governed by the Born--Infeld nonlinear theory is defined by the Lagrangian action density
\be\lb{1.1}
\mathcal{L}=b^2\left(1-\sqrt{1-\frac{2s}{b^2}}\right)+\frac{2}{(1+|\phi|^2)^2}D_{\mu}\phi\overline{D^{\mu}\phi}
-b^2\left(1-\sqrt{1-\frac{1}{b^2}\left(\frac{|\phi|^2-1}{|\phi|^2+1}\right)^2}\right),
\ee
where the quantity $s$ governing electromagnetism is as given in \eq{2.2}. We omit the Euler--Lagrange equations of \eq{1.1} but focus our attention on magnetic vortices and antivortices as
before such that the planar Hamiltonian energy density associated with \eq{1.1} reads
\be\lb{ax2.25}
\mathcal{H}=b^2\left(\sqrt{1+\frac{1}{b^2}{F}_{12}^2}-1\right)+\frac{2}{(1+|\phi|^2)^2}|D_i\phi|^2 +b^2\left(1-\sqrt{1-\frac{1}{b^2}\left(\frac{|\phi|^2-1}{|\phi|^2+1}\right)^2}\right).
\ee
It is interesting to note that  \eq{ax2.25} recovers \eq{ax2.9}, with \eq{ax2.14} and $\xi=1$, in the limit $|\phi|^2\to 1$. With \eq{ax2.25}, we rederive the associated Bogomol'nyi equations \cite{LY}
\bea
F_{12}\pm {\sqrt{1+\frac1{b^2}F_{12}^2}}\left(\frac{|\phi|^2-1}{|\phi|^2+1}\right)&=&0,\lb{1.4}\\
{\sqrt{1+\frac1{b^2}F_{12}^2}} \sqrt{1-\frac{1}{b^2}\left(\frac{|\phi|^2-1}{|\phi|^2+1}\right)^2} &=& 1,\lb{1.5}\\
D_1\phi\pm \ri D_2 \phi &=& 0,\lb{1.6}
\eea
for which \eq{1.4} and \eq{1.5} may be suppressed into the single equation
\be\lb{ax2.29}
F_{12}=\pm\left(\frac{1-|\phi|^2}{1+|\phi|^2}\right)\frac1{ \sqrt{1-\frac{1}{b^2}\left(\frac{|\phi|^2-1}{|\phi|^2+1}\right)^2}}.
\ee
In this situation, the equations govern a system of vortices and antivortices, realized  by the zeros and poles of $\phi$,
which are isolated and have integer multiplicities,  say $q_1,\dots, q_M$, and $p_1,\dots,p_N$, respectively, with repeated listings counting for multiplicities, and the  minimum energy is
\be
E=\int_{\bfR^2} {\cal H}\,\dd x=2\pi (M+N),
\ee
since now \eq{ax2.15} becomes
\be
{\cal H}\geq \pm (F_{12}+J_{12}),
\ee
with
\bea
\frac1{2\pi}\int_{\bfR^2}F_{12}\,\dd x&=&\pm(M-N),\\
\frac1{2\pi}\int_{\bfR^2} J_{12}\,\dd x&=&\pm 2N,
\eea
representing the first Chern class and the Thom class \cite{SSY,XY}, respectively, where $J_i$ is the current density defined by
\be\lb{ax2.33}
J_i=\frac{\ii}{|\phi|^2+1}(\phi \overline{D_i\phi}-\overline{\phi}D_i\phi),\quad i=1,2.
\ee
In fact, in the present planar setting, the Euler--Lagrange equations of \eq{1.1} are those of the Hamiltonian energy density \eq{ax2.25}:
\bea
D_i\left(\frac{D_i\phi}{(|\phi|^2+1)^2}\right)&=& -\frac{2|D_i\phi|^2 \phi}{(|\phi|^2+1)^3}+\frac{(|\phi|^2-1)\phi}{(|\phi|^2+1)^3\sqrt{1-\frac1{b^2}\left(\frac{|\phi|^2+1}{|\phi|^2+1}\right)^2}},\lb{ax2.34}\\
\pa_j\left(\frac{ F_{ij}}{\sqrt{1+\frac{F_{12}^2}{b^2}}}\right)&=&\frac{2\ri}{(|\phi|^2+1)^2}(\phi\overline{D_i\phi}-\overline{\phi}D_i\phi).\lb{ax2.35}
\eea
It may be shown that \eq{1.4}--\eq{1.6} imply \eq{ax2.34}--\eq{ax2.35}. Moreover, as before, using $u=\ln|\phi|^2$, it is seen that \eq{1.6} and \eq{ax2.29} may be recast into
the nonlinear elliptic equation
\be\lb{ax2.36}
\Delta u=\frac{2(\e^u-1)}{(\e^u+1)\sqrt{1-\frac1{b^2}\left(\frac{\e^u-1}{\e^u+1}\right)^2}}+4\pi\sum_{s=1}^M \delta_{q_s}(x)-4\pi\sum_{s=1}^N \delta_{p_s}(x),\quad x\in\bfR^2.
\ee

In \cite{Ybi} and \cite{LY},  the Bogomol'nyi vortex systems and vortex-antivortex systems are obtained through constructing solutions to the equations \eq{ax2.23} and \eq{ax2.36}, respectively.

In the subsequent sections, we shall consider the Born--Infeld theories built over the product Abelian gauge group $U(1)\times U(1)$ and concentrate on the systems of magnetic vortices and
 vortices and antivortices generated from various Bogomol'nyi equations. Such  structures allow us to skip their Minkowski spacetime formalisms and delve directly into
the associated planar-space settings governed by the underlying Hamiltonian energy density functions.

For clarity and convenience, the details of our work are outlined in the chart below:

\begin{center}
\begin{tabular}{@{}lllll@{}}
\toprule
Section & Gauge sectors & Species & Vortices & Antivortices \\
\midrule
3 & Born--Infeld / Born--Infeld & 2 & Yes & No \\
4 & Maxwell / Born--Infeld & 2 & Yes & No \\
5 & Born--Infeld / Born--Infeld & 2 & Yes & Yes \\
6--8 & Mixed types & 2 & Yes & Yes \\
9 & Thermodynamics (full plane)& 2 & Yes & Model-dependent \\
10 & Thermodynamics (bounded domain) & 2 &Yes & Model-dependent \\
\bottomrule
\end{tabular}
\end{center}

\section{Bogomol'nyi equations in two-species Born--Infeld  electrodynamics}\lb{sec3}

In this section we extend the single-species framework to a two-species setting governed by a product $U(1)\times U(1)$ gauge symmetry. The goal is to demonstrate that the Bogomol'nyi type equations persist even in the presence of multiple gauge fields and Higgs scalars coupled through distinct charge assignments. We begin by formulating the two-species Born--Infeld energy functional and then identify a distinguished class of potentials for which the energy admits an exact lower bound. The resulting first-order equations describe interacting vortices carrying flux in different gauge sectors. Beyond their intrinsic interest, these equations provide a natural arena for studying composite vortices and fractional flux phenomena arising from multi-component gauge interactions.

Use $\hat{A}_i$ and $\tilde{A}_i$ to denote two real-valued Abelian gauge fields and $\hat{F}_{ij}=\pa_{i}\hat{A}_j-\pa_j\hat{A}_i$ and $\tilde{F}_{ij}=\pa_i\tilde{A}_j-\pa_j\tilde{A}_i$ the respectively induced magnetic fields. The gauge-covariant derivatives applied to two complex scalar Higgs fields $\phi$ and $\psi$ carrying two pairs of charges, $(a,b)$ and $(c,d)$, are defined by
\be\lb{ax3.1}
D_i \phi =\pa_i \phi -\ri(a\hat{A}_i+b\tilde{A}_i)\phi, \quad \quad D_i \psi =\pa_i \psi -\ri(c\hat{A}_i+d\tilde{A}_i)\psi,\quad i=1,2,
\ee
subject to the nondegeneracy condition
\be\lb{a2.6}
ad-bc\neq0,
\ee
throughout this work. (The charge parameter $b$ here and in the sequel should not be confused with the Born parameter $b$ used in Section \ref{sec2}.)
Extending the single-species Born--Infeld theory consisting of \eq{2.1} and \eq{2.2} into a two-species theory with two Born parameters, say $b_1, b_2>0$, the planar-space Hamiltonian
energy density reads
\be
\mathcal{H}=b_1^2\left(\sqrt{1+\frac{1}{b_1^2}\hat{F}_{12}^2}-1\right)+b_2^2\left(\sqrt{1+\frac{1}{b_2^2}\tilde{F}_{12}^2}-1\right)
+\frac12|D_i\phi|^2+\frac12|D_i\psi|^2
+V,\lb{x2.6}
\ee
where $V=V(|\phi|^2,|\psi|^2)$ is the potential energy density function to be determined to achieve a Bogomol'nyi structure.
The Euler--Lagrange equations of \eq{x2.6} read
\bea
D^2_i\phi&=&2\frac{\pa V(|\phi|^2,|\psi|^2)}{\pa{\overline{\phi}}},\lb{c1}  \\
D^2_i\psi&=&2\frac{\pa V(|\phi|^2,|\psi|^2)}{\pa{\overline{\psi}}}, \lb{c2} \\
\pa_j\left(\frac{\hat{F}_{ij}}{\sqrt{1+\frac{1}{b_1^2}\hat{F}_{12}^2}}\right)&=&\frac{\ri}{2}(a[\phi\overline{D_i\phi}-\overline{\phi}D_i\phi])+ c[\psi\overline{D_i\psi}-\overline{\psi}D_i\psi]),\lb{c3}\\
\pa_j\left(\frac{\tilde{F}_{ij}}{\sqrt{1+\frac{1}{b_2^2}\tilde{F}_{12}^2}}\right)&=&\frac{\ri}{2}(b[\phi\overline{D_i\phi}-\overline{\phi}D_i\phi])+ d[\psi\overline{D_i\psi}-\overline{\psi}D_i\psi]).\lb{c4}
\eea
On the other hand, note that there hold the identities
\bea
|D_i \phi|^2&=&|D_1 \phi \pm \ri D_2 \phi|^2\pm \ri \left(\pa_1[\phi\overline{D_2 \phi}]-\pa_2[\phi\overline{D_1\phi}]\right)\pm (a\hat{F}_{12}+b\tilde{F}_{12})|\phi|^2, \lb{x2.7}\\
|D_i \psi|^2&=&|D_1 \psi \pm \ri D_2 \psi|^2\pm \ri \left(\pa_1[\psi\overline{D_2 \psi}]-\pa_2[\psi\overline{D_1\psi}]\right)\pm (c\hat{F}_{12}+d\tilde{F}_{12})|\psi|^2.\lb{x2.8}
\eea
Hence, in view of \eq{x2.7} and \eq{x2.8},  the energy density becomes
\bea\lb{x2.9}
\mathcal{H}
&=&\frac12\left(\hat{F}_{12}\pm \frac12 F\left(a(|\phi|^2-\xi)+c(|\psi|^2-\zeta)\right)\right)^2F^{-1}+\frac{b_1^2}{2}(FU-1)^2F^{-1}\nn \\
&&+\frac12\left(\tilde{F}_{12}\pm \frac12 G\left(b(|\phi|^2-\xi)+d(|\psi|^2-\zeta)\right)\right)^2G^{-1}+\frac{b_2^2}{2}(GW-1)^2 G^{-1}\nn \\
&&+\frac12|D_1 \phi \pm \ri D_2 \phi|^2+\frac12|D_1 \psi \pm \ri D_2 \psi|^2\pm \frac12(a\xi+c\zeta)\hat{F}_{12}\pm \frac12(b\xi+d\zeta)\tilde{F}_{12}\nn \\
&&\pm \frac{\ri}{2} \left(\pa_1[\phi\overline{D_2 \phi}]-\pa_2[\phi\overline{D_1\phi}]\right)\pm \frac{\ri}{2} \left(\pa_1[\psi\overline{D_2 \psi}]-\pa_2[\psi\overline{D_1\psi}]\right)\nn\\
&&-b_1^2+b_1^2U-b_2^2+b_2^2W+V(|\phi|^2,|\psi|^2),
\eea
where
\bea
F&=& {\sqrt{1+\frac1{b_1^2}\hat{F}_{12}^2}}, \quad G={\sqrt{1+\frac1{b_2^2}\tilde{F}_{12}^2}},\nn\\
U&=&\sqrt{1-\frac{(a(|\phi|^2-\xi)+c(|\psi|^2-\zeta))^2}{4b_1^2}}, \quad
W=\sqrt{1-\frac{(b(|\phi|^2-\xi)+d(|\psi|^2-\zeta))^2}{4b_2^2}}, \nn
\eea
and $\xi, \zeta>0$ are  constants giving rise to the energy scales of the spontaneously broken symmetries. Thus, if we choose
\bea
V(|\phi|^2,|\psi|^2)&=& b_1^2(1-U)+b_2^2(1-W)\nn \\
&=&b_1^2\left(1-\sqrt{1-\frac{(a(|\phi|^2-\xi)+c(|\psi|^2-\zeta))^2}{4b_1^2}}\right)\nn\\
&&+b_2^2\left(1-\sqrt{1-\frac{(b(|\phi|^2-\xi)+d(|\psi|^2-\zeta))^2}{4b_2^2}}\right),
\eea
which leads to
\be\lb{x2.10}
E=\int_{\bfR^2}\mathcal{H}\, \dd x\ge\pm \frac12(a\xi+c\zeta)\int_{\bfR^2}\hat{F}_{12}\, \dd x\pm\frac12(b\xi+d\zeta)\int_{\bfR^2}\tilde{F}_{12}\, \dd x,
\ee
which enables us to arrive at the Bogomol'nyi equations
\bea
\hat{F}_{12}\pm \frac12F[a(|\phi|^2-\xi)+c(|\psi|^2-\zeta)]&=&0,\lb{x2.11}\\
\tilde{F}_{12}\pm \frac12G[b(|\phi|^2-\xi)+d(|\psi|^2-\zeta)]&=&0,\lb{x2.12}\\
FU-1&=&0,\lb{x2.13}\\
GW-1&=&0,\lb{x2.14}\\
D_1\phi\pm \ri D_2 \phi &=& 0, \lb {x2.15}\\
D_1\psi\pm \ri D_2 \psi &=& 0,\lb{x2.16}
\eea
for the two-species Born--Infeld Higgs model \eq{x2.6}.
It is straightforward to examine that \eq{x2.11} and \eq{x2.13} and \eq{x2.12} and \eq{x2.14} give rise to
\bea
\hat{F}_{12}&=&\mp\frac{a(|\phi|^2-\xi)+c(|\psi|^2-\zeta)}{2\sqrt{1-\frac{(a(|\phi|^2-\xi)+c(|\psi|^2-\zeta))^2}{4b_1^2}}},\lb{ax3.18}\\
\tilde{F}_{12}&=&\mp\frac{b(|\phi|^2-\xi)+d(|\psi|^2-\zeta)}{2\sqrt{1-\frac{(b(|\phi|^2-\xi)+d(|\psi|^2-\zeta))^2}{4b_2^2}}},\lb{ax3.19}
\eea
respectively. It can be shown that the equations \eq{x2.11}-\eq{x2.16} or \eq{x2.15}--\eq{ax3.19} imply \eq{c1}-\eq{c4}.

Let the zeros of the complex scalar fields $\phi$ and $\psi$ be given by
\be\lb{4.0}
\phi: \left\{q'_1,\dots,q'_{M_1}\right\},\quad \psi:\left\{{q}''_1,\dots,{q}''_{M_2}\right\},
\ee
respectively. From the equations \eq{x2.15} and \eq{x2.16}, we obtain
\be\lb{4.1}
a\hat{F}_{12}+b\tilde{F}_{12}=\mp \frac12 \Delta \ln |\phi|^2, \quad c\hat{F}_{12}+d\tilde{F}_{12}=\mp \frac12 \Delta \ln |\psi|^2,
\ee
away from the zeros of $\phi$ and $\psi$. Let $u=\ln|\phi|^2$ and $v=\ln|\psi|^2$. Then, from \eq{ax3.18} and \eq{ax3.19}, we arrive at the coupled nonlinear elliptic equations
\bea
\Delta u&=&\left(\frac{a^2}{\sqrt{1-\frac{1}{4b_1^2} f_1^2}}+\frac{b^2}{\sqrt{1-\frac{1}{4b_2^2}f_2^2}}\right)(\mathrm{e}^u-\xi)\nn \\
&&+\left(\frac{ac}{\sqrt{1-\frac{1}{4b_1^2} f_1^2}}+\frac{bd}{\sqrt{1-\frac{1}{4b_2^2}f_2^2}}\right)(\mathrm{e}^v-\zeta)+4\pi\sum_{s=1}^{M_1} \delta_{q'_s},\lb{4.2}\\
\Delta v &=&\left(\frac{ac}{\sqrt{1-\frac{1}{4b_1^2} f_1^2}}+\frac{bd}{\sqrt{1-\frac{1}{4b_2^2}f_2^2}}\right)(\mathrm{e}^u-\xi)\nn \\
&&+\left(\frac{c^2}{\sqrt{1-\frac{1}{4b_1^2} f_1^2}}+\frac{d^2}{\sqrt{1-\frac{1}{4b_2^2}f_2^2}}\right)(\mathrm{e}^v-\zeta)+4\pi\sum_{s=1}^{M_2} \delta_{{q}''_s},\lb{4.3}
\eea
where
\be
f_1=f_1(u,v)=a(\e^u-\xi)+c(\e^v-\zeta),\quad f_2=f_2(u,v)=b(\e^u-\xi)+d(\e^v-\zeta).\lb{4.4}
\ee

We next calculate the magnetic fluxes and energy of the system of vortices represented by the sets of zeros of the complex scalar fields $\phi$ and $\psi$ given in \eq{4.0}.
In order to avoid inconvenience with the boundary terms, we consider a doubly-periodic domain, $\Omega$, to work with,  which is of independent interest because it describes Abrikosov's vortex
condensates \cite{Ab} for which $\Omega$ serves as a prototype lattice cell domain.
For this purpose, we use the background functions $u'_{0}$ and $u''_{0}$ (cf. \cite{Aubin}) such that
\be\lb{a4.4}
\Delta u'_{0}=-\frac{4 \pi M_1}{|\Om|}+4 \pi \sum_{s=1}^{M_1} \delta_{q'_{s}},\quad \Delta u''_{0}=-\frac{4 \pi M_2}{|\Om|}+4 \pi \sum_{s=1}^{M_2} \delta_{{q}''_{s}}.
\ee
Set $u=u'_{0}+w'$ and $v=u''_{0}+w''$. Then we can rewrite \eq{4.2} and \eq{4.3} into the following source-free equations
\bea
\Delta w'&=&\left(\frac{a^2}{\sqrt{1-\frac{1}{4b_1^2} {f}_1^2}}+\frac{b^2}{\sqrt{1-\frac{1}{4b_2^2}{f}_2^2}}\right)(\mathrm{e}^{u'_{0}+w'}-\xi)\nn \\
&&+\left(\frac{ac}{\sqrt{1-\frac{1}{4b_1^2}{f}_1^2}}+\frac{bd}{\sqrt{1-\frac{1}{4b_2^2}{f}_2^2}}\right)(\mathrm{e}^{u''_{0}+w''}-\zeta)+\frac{4 \pi M_1}{|\Om|},\lb{4.4a}\\
\Delta w''&=&\left(\frac{ac}{\sqrt{1-\frac{1}{4b_1^2}{f}_1^2}}
+\frac{bd}{\sqrt{1-\frac{1}{4b_2^2}{f}_2^2}}\right)(\mathrm{e}^{u'_{0}+w'}-\xi)\nn \\
&&+\left(\frac{c^2}{\sqrt{1-\frac{1}{4b_1^2}{f}_1^2}}+\frac{d^2}{\sqrt{1-\frac{1}{4b_2^2}{f}_2^2}}\right)(\mathrm{e}^{u''_{0}+w''}-\zeta)+\frac{4\pi M_2}{|\Om|},\lb{4.5}
\eea
where the quantities $f_1$ and $f_2$ are as defined in \eq{4.4} but updated:
\bea
{f}_1&=&f_1(u'_0+w',u_0''+w'')=a(\e^{u'_0+w'}-\xi)+c(\e^{u''_0+w''}-\zeta),\lb{e3}\\
{f}_2&=&f_2(u'_0+w',u''_0+w'')=b(\e^{u'_0+w'}-\xi)+d(\e^{u''_0+w''}-\zeta).\lb{e4}
\eea
Integrating \eq{4.4a} and \eq{4.5}, we obtain the quantized integrals
\bea
\int_{\Om}\frac{a(\e^{u'_0+w'}-\xi)+c(\e^{u_0''+w''}-\zeta)}{\sqrt{1-\frac{1}{4b_1^2} f_1^2}}\dd x=\frac{4\pi (b M_2-d M_1)}{ad-bc},\lb{4.6}\\
\int_{\Om}\frac{b(\e^{u'_0+w'}-\xi)+d(\e^{u_0''+w''}-\zeta))}{\sqrt{1-\frac{1}{4b_2^2} f_2^2}}\dd x=\frac{4\pi (c M_1-a M_2)}{ad-bc}. \lb{4.7}
\eea
Inserting \eq{4.6} and \eq{4.7} into  of \eq{ax3.18} and \eq{ax3.19}, we have the fluxes
\bea
\hat{\Phi}&=&\int_\Om\hat{F}_{12}\, \dd x=\pm \frac{2\pi(d M_1-b M_2)}{ad-bc},\lb{4.10a} \\
\tilde{\Phi}&=&\int_\Om\tilde{F}_{12}\, \dd x=\pm \frac{2\pi(a M_2-c M_1)}{ad-bc}.\lb{4.10b}
\eea
In view of \eq{4.10a}, \eq{4.10b}, and \eq{x2.10}, we obtain the energy
\bea
E&=&\int_\Om \mathcal{H}\, \dd x=\pm\frac12 \left((a\xi+c\zeta)\hat{\Phi}+(b\xi+d\zeta)\tilde{\Phi}\right)\nn\\
&=&\pi \left(\xi M_1+\zeta M_2\right),\lb{4.10c}
\eea
which is elegantly expressed in terms of the vortex numbers $M_1$ and $M_2$. Note that this result is independent of the domain $\Om$ chosen and the charge parameters $a,b,c,d$.

\section{Bogomol'nyi equations in two-species Maxwell and Born--Infeld electrodynamics}\lb{sec4}

The purpose of this section is to investigate hybrid gauge theories in which one gauge sector is governed by the classical Maxwell action while the other retains the Born--Infeld nonlinearity. Such mixed models interpolate between linear and nonlinear electrodynamics and allow us to isolate the effects of the Born--Infeld regularization within a multi-species context. We show that, despite this asymmetry, the system still admits a Bogomol'nyi structure under suitable choices of the potential. The resulting equations reveal how the Maxwell and Born--Infeld vortices coexist and interact, and they prepare the ground for the vortex--antivortex extensions considered later.

The planar energy density in this situation reads
\be\lb{ax4.1}
\mathcal{H}= \frac{1}{2}\hat{F}_{12}^2+b_1^2\left(\sqrt{1+\frac{1}{b_1^2}\tilde{F}_{12}^2}-1\right)+\frac12|D_i\phi|^2+\frac12|D_i\psi|^2 +V,
\ee
where $D_i\phi$ and $D_i\psi$ are as defined in \eq{ax3.1}.
The Euler--Lagrange equations are:
\bea
D^2_i\phi&=&2\frac{\pa V}{\pa\overline{\phi}}, \lb{x2.18} \\
D^2_i\psi&=&2\frac{\pa V}{\pa\overline{\psi}}, \lb{x2.19} \\
\pa_j{\hat{F}_{ij}}&=&\frac{\ri}{2}(a[\phi\overline{D_i\phi}-\overline{\phi}D_i\phi])+ c[\psi\overline{D_i\psi}-\overline{\psi}D_i\psi]),\lb{x2.20}\\
\pa_j\left(\frac{\tilde{F}_{ij}}{\sqrt{1+\frac{1}{b_1^2}\tilde{F}_{12}^2}}\right)&=&\frac{\ri}{2}(b[\phi\overline{D_i\phi}-\overline{\phi}D_i\phi])+ d[\psi\overline{D_i\psi}-\overline{\psi}D_i\psi]).\lb{x2.21}
\eea
Applying the identities \eq{x2.7} and \eq{x2.8}, we rewrite \eq{ax4.1}  as
\bea
\mathcal{H}
&=&\frac{1}{2}\bigg(\hat{F}_{12}\pm \frac{1}{2}\left(a(|\phi|^2-\xi)+c(|\psi|^2-\zeta)\right)\bigg)^2\nn \\
&&+\frac{1}{2}\bigg(\tilde{F}_{12}\pm \frac{1}{2}F\left(b(|\phi|^2-\xi)+d(|\psi|^2-\zeta)\right)\bigg)^2F^{-1}+\frac{b_1^2}{2}(FU-1)^2 F^{-1}\nn \\
&&+\frac12|D_1 \phi \pm \ri D_2 \phi|^2+\frac12|D_1 \psi \pm \ri D_2 \psi|^2\pm \frac12(a\xi+c\zeta)\hat{F}_{12}\pm \frac12(b\xi+d\zeta)\tilde{F}_{12}\nn \\
&&\pm \frac{\ri}{2}\left(\pa_1[\phi\overline{D_2 \phi}]-\pa_2[\phi\overline{D_1\phi}]\right)\pm \frac{\ri}{2} \left(\pa_1[\psi\overline{D_2 \psi}]-\pa_2[\psi\overline{D_1\psi}]\right),\lb{x2.23}
\eea
where $F$ and $U$ are
\be
F={\sqrt{1+\frac1{b_1^2}\tilde{F}_{12}^2}}, \quad U= \sqrt{1-\frac{1}{4b_1^2}(b[|\phi|^2-\xi]+d[|\psi|^2-\zeta])^2},
\ee
and the potential density is chosen to be
\be\lb{ax4.8}
V=\frac{1}{8}\left(a[|\phi|^2-\xi]+c[|\psi|^2-\zeta]\right)^2
+b_1^2\left(1-\sqrt{1-\frac{1}{4b_1^2}(b[|\phi|^2-\xi]+d[|\psi|^2-\zeta])^2}\right).
\ee
We observe that the first term in \eq{ax4.8} corresponds to the potential density of the classical Abelian Higgs (Maxwell) theory and the second term is that of the Born--Infeld theory.
As a result, the total energy has the lower bound
\be\lb{ax4.9}
E=\int\mathcal{H}\, \dd x\ge\pm \frac12(a\xi+c\zeta)\int\hat{F}_{12}\, \dd x\pm \frac12(b\xi+d\zeta)\int\tilde{F}_{12}\, \dd x,
\ee
integrated over the full domain of interest.
This lower bound is attained if the squares on the right-hand side of \eq{x2.23} all vanish:
\bea
\hat{F}_{12}\pm \frac{1}{2}(a[|\phi|^2-\xi]+c[|\psi|^2-\zeta])&=&0,\lb{x2.24}\\
\tilde{F}_{12}\pm \frac{1}{2} F(b[|\phi|^2-\xi]+d[|\psi|^2-\zeta])&=&0,\lb{x2.25}\\
FU-1&=&0,\lb{x2.26}\\
D_1\phi\pm \ri D_2 \phi &=& 0, \lb {x2.27}\\
D_1\psi\pm \ri D_2 \psi &=& 0. \lb{x2.28}
\eea
These are the Bogomol'nyi equations of the model \eq{ax4.1}.
It is clear that \eq{x2.25} and \eq{x2.26} may be combined into one equation:
\be\lb{ax4.15}
\tilde{F}_{12}=\mp\frac{b(|\phi|^2-\xi)+d(|\psi|^2-\zeta)}{2\sqrt{1-\frac{1}{4b_1^2}(b[|\phi|^2-\xi]+d[|\psi|^2-\zeta])^2}}.
\ee
Thus, with \eq{4.0}, \eq{4.1}, and $u=\ln|\phi|^2, v=\ln|\psi|^2$, we obtain the Bogomol'nyi associated vortex equations
\bea
\Delta u&=&\left({a^2}+\frac{b^2}{\sqrt{1-\frac{1}{4b_1^2}f_2^2}}\right)(\mathrm{e}^u-\xi)
+\left({ac}+\frac{bd}{\sqrt{1-\frac{1}{4b_1^2}f_2^2}}\right)(\mathrm{e}^v-\zeta)+4\pi\sum_{s=1}^{M_1} \delta_{q'_s},\nn\\
\lb{4.11}\\
\Delta v &=&\left({ac}+\frac{bd}{\sqrt{1-\frac{1}{4b_1^2}f_2^2}}\right)(\mathrm{e}^u-\xi)
+\left({c^2}+\frac{d^2}{\sqrt{1-\frac{1}{4b_1^2}f_2^2}}\right)(\mathrm{e}^v-\zeta)+4\pi\sum_{s=1}^{M_2} \delta_{{q}''_s},\nn\\
\lb{4.12}
\eea
where the quantity $f_2$ is  as defined in \eq{4.4}. Using the same method as in Section \ref{sec3}, we can integrate \eq{4.11} and \eq{4.12} to get the equations
\bea
a\int\left(a(\e^u-\xi)+c(\e^v-\zeta)\right)\dd x+b\int\left(\frac{b(\e^u-\xi)+d(\e^v-\zeta)}{\sqrt{1-\frac1{4b_1^2}f_2^2}}\right)\,\dd x&=&-4\pi M_1,\quad \\
c\int\left(a(\e^u-\xi)+c(\e^v-\zeta)\right)\dd x+d\int\left(\frac{b(\e^u-\xi)+d(\e^v-\zeta)}{\sqrt{1-\frac1{4b_1^2}f_2^2}}\right)\,\dd x&=&-4\pi M_2.\quad
\eea
Solving these equations, we obtain
\bea
\int\left(a(\e^u-\xi)+c(\e^v-\zeta)\right)\dd x&=&\frac{4\pi(bM_2-dM_1)}{ad-bc},\lb{ax4.20}\\
\int\left(\frac{b(\e^u-\xi)+d(\e^v-\zeta)}{\sqrt{1-\frac1{4b_1^2}f_2^2}}\right)\,\dd x&=&\frac{4\pi(cM_1-aM_2)}{ad-bc},\lb{ax4.21}
\eea
which are similar to \eq{4.6} and \eq{4.7}. In view of \eq{ax4.20} and \eq{ax4.21}, we see that the vorticity field equations \eq{x2.24} and \eq{ax4.15} render us the magnetic fluxes
\bea
\hat{\Phi}&=&\int \hat{F}_{12}\,\dd x=\pm\frac{2\pi(dM_1-bM_2)}{ad-bc},\lb{ax4.22}\\
\tilde{\Phi}&=&\int\tilde{F}_{12}\,\dd x=\pm\frac{2\pi(aM_2-cM_1)}{ad-bc},\lb{ax4.23}
\eea
which are formally identical to \eq{4.10a} and \eq{4.10b}. Hence, in view of these results and \eq{ax4.9}, we arrive at the same formula  \eq{4.10c} for the total quantized minimum energy
of the two-species vortex system here.

\section{Systems of vortices and antivortices in two-species  Born--Infeld model}\lb{sec5}

This section generalizes the two-species Born--Infeld vortex model to include antivortices through the introduction of gauged harmonic map type kinetic terms. The presence of both vortices and antivortices significantly enriches the topological and analytical structure of the theory, leading to competing contributions to magnetic flux and energy. We derive the corresponding Bogomol'nyi equations and identify the topological invariants governing the system. These results highlight the symmetry between vortices and antivortices and lay the foundation for the thermodynamic analysis of neutral and mixed configurations in later sections.

Combining and extending the theories \eq{ax2.25} and \eq{x2.6}, we consider the two-species Born--Infeld model whose
 static energy density  is given by
\be
\mathcal{H}=b_1^2\left(\sqrt{1+\frac{1}{b_1^2}\hat{F}_{12}^2}-1\right)+b_2^2\left(\sqrt{1+\frac{1}{b_2^2}\tilde{F}_{12}^2}-1\right)
+\frac{2|D_i\phi|^2 }{(1+|\phi|^2)^2}
+\frac{2|D_i\psi|^2}{(1+|\psi|^2)^2}+V,\lb{3.10}
\ee
such that both vortices and antivortices are accommodated.
The Euler--Lagrange equations of \eq{3.10} are
\bea
D_i\left(\frac{D_i\phi}{(|\phi|^2+1)^2}\right)&=& -\frac{2|D_i\phi|^2 \phi}{(|\phi|^2+1)^3}+\frac12\frac{\pa V(|\phi|^2,|\psi|^2)}{\pa \overline{\phi}},\lb{d1}\\
D_i\left(\frac{D_i\psi}{(|\psi|^2+1)^2}\right)&=& -\frac{2|D_i\psi|^2 \psi}{(|\psi|^2+1)^3}+\frac12\frac{\pa V(|\phi|^2,|\psi|^2)}{\pa \overline{\psi}},\lb{d2}\\
\pa_j\left(\frac{ \hat{F}_{ij}}{\sqrt{1+\frac{\hat{F}_{12}^2}{b_1^2}}}\right)&=&\frac{2 \ri(a[\phi\overline{D_i\phi}-\overline{\phi}D_i\phi])}{(1+|\phi|^2)^2}+\frac{2 \ri(c[\psi\overline{D_i\psi}-\overline{\psi}D_i\psi])}{(1+|\psi|^2)^2},
\lb{d3}\\
\pa_j\left(\frac{ \tilde{F}_{ij}}{\sqrt{1+\frac{\tilde{F}_{12}^2}{b_2^2}}}\right)&=&\frac{2 \ri(b[\phi\overline{D_i\phi}-\overline{\phi}D_i\phi])}{(1+|\phi|^2)^2}+\frac{2 \ri(d[\psi\overline{D_i\psi}-\overline{\psi}D_i\psi])}{(1+|\psi|^2)^2}.
\lb{d4}
\eea
In order to unveil a Bogomol'nyi structure, we introduce two current densities
\be
J_i(\phi)=\frac{\ri}{1+|\phi|^2}(\phi \overline{D_i\phi}-\overline{\phi}D_i\phi),\quad
J_i(\psi)=\frac{\ri}{1+|\psi|^2}(\psi\overline{D_i \psi}-\overline{\psi}D_i \psi),\quad i=1,2,\lb{3.12}
\ee
resulting in the relations
\bea
J_{12}(\phi)&=&\frac{-2|\phi|^2}{1+|\phi|^2}(a \hat{F}_{12}+b\tilde{F}_{12})+\frac{2\ri}{(1+|\phi|^2)^2}(D_1 \phi\overline{D_2 \phi}-\overline{D_1\phi}D_2 \phi),\lb{3.13} \\
J_{12}(\psi)&=&\frac{-2|\psi|^2}{1+|\psi|^2}(c \hat{F}_{12}+d\tilde{F}_{12})+\frac{2\ri}{(1+|\psi|^2)^2}(D_1 \psi\overline{D_2 \psi}-\overline{D_1\psi}D_2 \psi).\lb{3.14}
\eea
For convenience, we shall apply the notation
\bea
&&F= {\sqrt{1+\frac1{b_1^2}\hat{F}_{12}^2}}, \quad G= {\sqrt{1+\frac1{b_2^2}\tilde{F}_{12}^2}},\\
&&U= \sqrt{1-\frac{1}{b_1^2}\left(a\frac{|\phi|^2-1}{|\phi|^2+1}+c\frac{|\psi|^2-1}{|\psi|^2+1}\right)^2}, \\
&& W=\sqrt{1-\frac{1}{b_2^2}\left(b\frac{|\phi|^2-1}{|\phi|^2+1}+d\frac{|\psi|^2-1}{|\psi|^2+1}\right)^2}.
\eea
Thus, using \eq{ax2.11} for $\phi$ and a similar expression for $\psi$,  we can rewrite  \eq{3.10} as
\bea\lb{3.15}
\mathcal{H}
&=& \frac12\left(\hat{F}_{12}\pm F\left(a\frac{|\phi|^2-1}{|\phi|^2+1}+c\frac{|\psi|^2-1}{|\psi|^2+1}\right)\right)^2F^{-1}+\frac{b_1^2}{2}(FU-1)^2F^{-1}\nn \\
&&+\frac12\left(\tilde{F}_{12}\pm G\left(b\frac{|\phi|^2-1}{|\phi|^2+1}+d\frac{|\psi|^2-1}{|\psi|^2+1}\right)\right)^2 G^{-1}+\frac{b_2^2}{2}(GW-1)^2 G^{-1} \nn \\
&&+\frac{2}{(1+|\phi|^2)^2}|D_1 \phi \pm \ri D_2 \phi|^2+\frac{2}{(1+|\psi|^2)^2}|D_1 \psi \pm \ri D_2 \psi|^2 \nn \\
&&\pm \left(\frac{-2|\phi|^2}{1+|\phi|^2}(a\hat{F}_{12}+b\tilde{F}_{12})\right)\pm \frac{2\ri}{(1+|\phi|^2)^2}(D_1\phi\overline{D_2 \phi}-\overline{D_1\phi}D_2 \phi) \nn \\
&&\pm \left(\frac{-2|\psi|^2}{1+|\psi|^2}(c\hat{F}_{12}+d\tilde{F}_{12})\right)\pm \frac{2\ri}{(1+|\psi|^2)^2}(D_1\psi\overline{D_2 \psi}-\overline{D_1\psi}D_2 \psi)\nn \\
&&\pm(a\hat{F}_{12}+b\tilde{F}_{12})\pm(c\hat{F}_{12}+d\tilde{F}_{12})-b_1^2(1-U)-b_2^2(1-W)
+V,\lb{2.4}
\eea
which leads us to choose the potential density $V$ to be
\bea
V&=& b_1^2(1-U)+b_2^2(1-W)\nn \\
&=&b_1^2\left(1-\sqrt{1-\frac{1}{b_1^2}\left(a\frac{|\phi|^2-1}{|\phi|^2+1}+c\frac{|\psi|^2-1}{|\psi|^2+1}\right)^2}\right)\nn\\
&&+b_2^2\left(1-\sqrt{1-\frac{1}{b_2^2}\left(b\frac{|\phi|^2-1}{|\phi|^2+1}+d\frac{|\psi|^2-1}{|\psi|^2+1}\right)^2}\right).\lb{ax5.10}
\eea
Using \eq{3.13}, \eq{3.14}, and \eq{ax5.10}  in \eq{3.15}, we arrive at the expression
\bea
\mathcal{H}&=&\frac12\left(\hat{F}_{12}\pm F\left(a\frac{|\phi|^2-1}{|\phi|^2+1}+c\frac{|\psi|^2-1}{|\psi|^2+1}\right)\right)^2F^{-1}+\frac{b_1^2}{2}(FU-1)^2F^{-1}\nn \\
&&+\frac12\left(\tilde{F}_{12}\pm G\left(b\frac{|\phi|^2-1}{|\phi|^2+1}+d\frac{|\psi|^2-1}{|\psi|^2+1}\right)\right)^2 G^{-1}+\frac{b_2^2}{2}(GW-1)^2 G^{-1} \nn \\
&&+\frac{2}{(1+|\phi|^2)^2}|D_1 \phi \pm \ri D_2 \phi|^2+\frac{2}{(1+|\psi|^2)^2}|D_1 \psi \pm \ri D_2 \psi|^2 \nn \\
&&\pm(a+c)\hat{F}_{12} \pm(b+d)\tilde{F}_{12}\pm J_{12}(\phi)\pm J_{12}(\psi).\lb{3.16}
\eea
Therefore the energy enjoys the topological  lower bound
\bea\lb{3.17}
E&=&\int_{\bfR^2}\mathcal{H}\,\dd x\nn\\
&\ge&\pm (a+c)\int_{\bfR^2}\hat{F}_{12}\,\dd x\pm (b+d)\int_{\bfR^2}\tilde{F}_{12}\,\dd x\pm\int_{\bfR^2}  J_{12}(\phi)\,\dd x\pm \int_{\bfR^2} J_{12}(\psi)\,\dd x,\quad\quad
\eea
which is attained  if all the quadratic terms in \eq{3.16} vanish,
\bea
\hat{F}_{12}\pm F\left(a\frac{|\phi|^2-1}{|\phi|^2+1}+c\frac{|\psi|^2-1}{|\psi|^2+1}\right)&=&0,\lb{3.18}\\
\tilde{F}_{12}\pm G\left(b\frac{|\phi|^2-1}{|\phi|^2+1}+d\frac{|\psi|^2-1}{|\psi|^2+1}\right)&=&0,\lb{3.19}\\
FU-1&=&0,\lb{3.20}\\
GW-1&=&0,\lb{3.21}\\
D_1\phi\pm \ri D_2 \phi &=& 0, \lb {3.22}\\
D_1\psi\pm \ri D_2 \psi &=& 0, \lb{3.23}
\eea
which are the Bogomol'nyi equations of the model \eq{3.10}. As before, we may combine \eq{3.18} and \eq{3.20} and \eq{3.19} and \eq{3.21} into the following two vorticity field equations,
\bea
\hat{F}_{12}&=&\mp\frac{a\frac{|\phi|^2-1}{|\phi|^2+1}+c\frac{|\psi|^2-1}{|\psi|^2+1}}{\sqrt{1-\frac{1}{b_1^2}\left(a\frac{|\phi|^2-1}{|\phi|^2+1}+c\frac{|\psi|^2-1}{|\psi|^2+1}\right)^2}},\lb{ax5.22}\\
\tilde{F}_{12}&=&\mp \frac{b\frac{|\phi|^2-1}{|\phi|^2+1}+d\frac{|\psi|^2-1}{|\psi|^2+1}}{\sqrt{1-\frac{1}{b_2^2}\left(b\frac{|\phi|^2-1}{|\phi|^2+1}+d\frac{|\psi|^2-1}{|\psi|^2+1}\right)^2}}.\lb{ax5.23}
\eea
Therefore, with the zeros and poles assigned to the complex scalar fields $\phi$ and $\psi$, respectively, by
\be
\phi: \left\{q'_1,\dots,q'_{M_1}\right\},\quad \left\{p'_1,\dots,p'_{N_1}\right\}; \quad \psi: \left\{{q}''_1,\dots,{q}''_{M_2}\right\},\quad
\left\{{p}''_1,\dots,{p}''_{N_2}\right\},\lb{5.2}
\ee
the relations in \eq{4.1}, and the substitutions of the variables $u=\ln|\phi|^2$ and $v=\ln|\psi|^2$, we obtain the associated Bogomol'nyi vortex-antivortex equations
\bea
\Delta u&=&2\left(\frac{a^2}{\sqrt{1-\frac{1}{b_1^2} f_3^2}}+\frac{b^2}{\sqrt{1-\frac{1}{b_2^2}f_4^2}}\right)\left(\frac{\e^u-1}{\e^u+1}\right)\nn \\
&&+2\left(\frac{ac}{\sqrt{1-\frac{1}{b_1^2} f_3^2}}+\frac{bd}{\sqrt{1-\frac{1}{b_2^2}f_4^2}}\right)\left(\frac{\e^v-1}{\e^v+1}\right)+4\pi\sum_{s=1}^{M_1} \delta_{q'_s}-4\pi\sum_{s=1}^{N_1} \delta_{p'_s},\quad \lb{5.5}\\
\Delta v &=&2\left(\frac{ac}{\sqrt{1-\frac{1}{b_1^2} f_3^2}}+\frac{bd}{\sqrt{1-\frac{1}{b_2^2}f_4^2}}\right)\left(\frac{\e^u-1}{\e^u+1}\right)\nn \\
&&+2\left(\frac{c^2}{\sqrt{1-\frac{1}{b_1^2} f_3^2}}+\frac{d^2}{\sqrt{1-\frac{1}{b_2^2}f_4^2}}\right)\left(\frac{\mathrm{e}^v-1}{\e^v+1}\right)+4\pi\sum_{s=1}^{M_2} \delta_{{q}''_s}-4\pi\sum_{s=1}^{N_2} \delta_{{p}''_s},\quad \lb{5.6}
\eea
where  the quantities $f_3=f_3(u,v)$ and $f_4=f_4(u,v)$ are defined by
\be
f_3=a\frac{\e^u-1}{\e^u+1}+c\frac{\e^v-1}{\e^v+1},\quad f_4=b\frac{\e^u-1}{\e^u+1}+d\frac{\e^v-1}{\e^v+1}.\lb{ax5.27}
\ee
As in Section \ref{sec3}, we can integrate \eq{5.5} and \eq{5.6} to get
\bea
a\int\frac{f_3}{\sqrt{1-\frac1{b_1^2} f_3^2}}\dd x+b\int\frac{f_4}{\sqrt{1-\frac1{b_2^2} f_4^2}}\dd x&=& -2\pi(M_1-N_1),\\
c\int\frac{f_3}{\sqrt{1-\frac1{b_1^2} f_3^2}}\dd x+d\int\frac{f_4}{\sqrt{1-\frac1{b_2^2} f_4^2}}\dd x&=&-2\pi(M_2-N_2).
\eea
Solving these equations, we obtain
\bea
\int\frac{f_3}{\sqrt{1-\frac1{b_1^2} f_3^2}}\dd x&=&\frac{2\pi(b(M_2-N_2)-d(M_1-N_1))}{ad-bc},\lb{5.7}\\
\int\frac{f_4}{\sqrt{1-\frac1{b_2^2} f_4^2}}\dd x&=&\frac{2\pi(c(M_1-N_1)-a(M_2-N_2))}{ad-bc}.\lb{5.8}
\eea
Substituting these results into \eq{ax5.22} and \eq{ax5.23}, we get the magnetic fluxes:
\bea
\hat{\Phi}&=&\int\hat{F}_{12}\,\dd x=\pm\frac{2\pi(d(M_1-N_1)-b(M_2-N_2))}{ad-bc},\lb{ax5.32}\\
\tilde{\Phi}&=&\int\tilde{F}_{12}\,\dd x=\pm\frac{2\pi(a(M_2-N_2)-c(M_1-N_1))}{ad-bc},\lb{ax5.33}
\eea
which extend the pairs \eq{4.10a}, \eq{4.10b} and \eq{ax4.22}, \eq{ax4.23} for the systems of vortices, without antivortices.
Besides, from \cite{SSY,XY}, we have
\be
\int J_{12}(\phi)\,\dd x=\pm 4\pi N_1,\quad \int J_{12}(\psi)\,\dd x=\pm 4\pi N_2.\lb{ax5.34}
\ee
Therefore, using \eq{ax5.32}--\eq{ax5.34} in \eq{3.17}, we find
\be\lb{ax5.35}
E=2\pi (M_1+M_2+N_1+N_2).
\ee

\section{Systems of vortices and antivortices governed by two-species Maxwell and Born--Infeld electrodynamics}\lb{sec6}

Here we combine the features of the previous two sections by considering vortex--antivortex systems governed simultaneously by the Maxwell and Born--Infeld electrodynamics. The objective is to understand how linear and nonlinear gauge dynamics influence the balance between vortices and antivortices and the resulting topological energy bounds. We show that the Bogomol'nyi framework remains applicable and yields explicit first-order equations and flux quantization formulas. This mixed setting provides additional flexibility and physical insight, especially in comparing purely Born--Infeld systems with their Maxwell counterparts.

The static energy density is
\be\lb{3.25}
\mathcal{H}=\frac{1}{2}\hat{F}_{12}^2+b_1^2\left(\sqrt{1+\frac{1}{b_1^2}\tilde{F}_{12}^2}-1\right)+\frac{2}{(1+|\phi|^2)^2}|D_i\phi|^2
+\frac{2}{(1+|\psi|^2)^2}|D_i\psi|^2
+V.
\ee
The Euler--Lagrange equations  of \eq{3.25} are
\bea
D_i\left(\frac{D_i\phi}{(|\phi|^2+1)^2}\right)&=& -\frac{2|D_i\phi|^2 \phi}{(|\phi|^2+1)^3}+\frac12\frac{\pa V}{\pa \overline{\phi}},\lb{d1}\\
D_i\left(\frac{D_i\psi}{(|\psi|^2+1)^2}\right)&=& -\frac{2|D_i\psi|^2 \psi}{(|\psi|^2+1)^3}+\frac12\frac{\pa V}{\pa \overline{\psi}},\lb{d2}\\
\pa_j\hat{F}_{ij}&=&\frac{2 \ri(a[\phi\overline{D_i\phi}-\overline{\phi}D_i\phi])}{(1+|\phi|^2)^2}+\frac{2 \ri(c[\psi\overline{D_i\psi}-\overline{\psi}D_i\psi])}{(1+|\psi|^2)^2},\lb{d3}\\
\pa_j\left(\frac{ \tilde{F}_{ij}}{\sqrt{1+\frac{\tilde{F}_{12}^2}{b_1^2}}}\right)&=&\frac{2 \ri(b[\phi\overline{D_i\phi}-\overline{\phi}D_i\phi])}{(1+|\phi|^2)^2}+\frac{2 \ri(d[\psi\overline{D_i\psi}-\overline{\psi}D_i\psi])}{(1+|\psi|^2)^2}.
\lb{d4}
\eea
On the other hand, using \eq{ax2.11} for $\phi$ and a similar expression for $\psi$, we can rewrite the Hamiltonian density \eq{3.25} as
\bea
\mathcal{H}
&=&\frac{1}{2}\left(\hat{F}_{12}\pm\left(a\frac{|\phi|^2-1}{|\phi|^2+1}+c\frac{|\psi|^2-1}{|\psi|^2+1}\right)\right)^2\nn \\
&&+\frac{1}{2}\left(\tilde{F}_{12}\pm F\left(b\frac{|\phi|^2-1}{|\phi|^2+1}+d\frac{|\psi|^2-1}{|\psi|^2+1}\right)\right)^2 F^{-1}+\frac{b_1^2}{2}(FU-1)^2F^{-1}\nn \\
&&+\frac{2|D_1 \phi \pm \ri D_2 \phi|^2}{(1+|\phi|^2)^2}+\frac{2|D_1 \psi \pm \ri D_2 \psi|^2 }{(1+|\psi|^2)^2}\nn \\
&&\pm \left(\frac{-2|\phi|^2}{1+|\phi|^2}(a\hat{F}_{12}+b\tilde{F}_{12})\right)\pm \frac{2\ri(D_1\phi\overline{D_2 \phi}-\overline{D_1\phi}D_2 \phi)}{(1+|\phi|^2)^2} \nn \\
&&\pm \left(\frac{-2|\psi|^2}{1+|\psi|^2}(c\hat{F}_{12}+d\tilde{F}_{12})\right)\pm \frac{2\ri(D_1\psi\overline{D_2 \psi}-\overline{D_1\psi}D_2 \psi)}{(1+|\psi|^2)^2}\nn \\
&&\pm(a\hat{F}_{12}+b\tilde{F}_{12})\pm(c\hat{F}_{12}+d\tilde{F}_{12})\nn\\
&&-\frac{1}{2}
\left(a\frac{|\phi|^2-1}{|\phi|^2+1}+c\frac{|\psi|^2-1}{|\psi|^2+1}\right)^2
-b_1^2(1-U)+V, \lb{3.26}
\eea
where $F$ and $U$ are
\be
F={\sqrt{1+\frac1{b_1^2}\tilde{F}_{12}^2}}, \quad U= \sqrt{1-\frac{1}{b_1^2}\left(b\frac{|\phi|^2-1}{|\phi|^2+1}+d\frac{|\psi|^2-1}{|\psi|^2+1}\right)^2}.
\ee
Thus we choose the potential density $V=V(|\phi|^2,|\psi|^2)$ to be
\be\lb{ax6.8}
V=\frac{1}{2}\left(a\frac{|\phi|^2-1}{|\phi|^2+1}+c\frac{|\psi|^2-1}{|\psi|^2+1}\right)^2 +b_1^2\left(1-\sqrt{1-\frac{1}{b_1^2}\left(b\frac{|\phi|^2-1}{|\phi|^2+1}+d\frac{|\psi|^2-1}{|\psi|^2+1}\right)^2}\right).
\ee
Using \eq{3.13}, \eq{3.14}, and \eq{ax6.8}  in \eq{3.26}, we arrive at
\bea
\mathcal{H}&=&\frac{1}{2}\left(\hat{F}_{12}\pm\left(a\frac{|\phi|^2-1}{|\phi|^2+1}+c\frac{|\psi|^2-1}{|\psi|^2+1}\right)\right)^2\nn \\
&&+\frac{1}{2}\left(\tilde{F}_{12}\pm F\left(b\frac{|\phi|^2-1}{|\phi|^2+1}+d\frac{|\psi|^2-1}{|\psi|^2+1}\right)\right)^2 F^{-1}+\frac{b_1^2}{2}(FU-1)^2F^{-1}\nn \\
&&+\frac{2}{(1+|\phi|^2)^2}|D_1 \phi \pm \ri D_2 \phi|^2+\frac{2}{(1+|\psi|^2)^2}|D_1 \psi \pm \ri D_2 \psi|^2 \nn \\
&&\pm(a+c)\hat{F}_{12}\pm(b+d)\tilde{F}_{12}\pm J_{12}(\phi)\pm J_{12}(\psi),\lb{x3.26}
\eea
which is similar to \eq{3.16}. Hence we get \eq{3.17}
whose saturation leads to the Bogomol'nyi equations
\bea
\hat{F}_{12}\pm \left(a\frac{|\phi|^2-1}{|\phi|^2+1}+c\frac{|\psi|^2-1}{|\psi|^2+1}\right)&=&0,\lb{3.28}\\
\tilde{F}_{12}\pm  F\left(b\frac{|\phi|^2-1}{|\phi|^2+1}+d\frac{|\psi|^2-1}{|\psi|^2+1}\right)&=&0,\lb{3.29}\\
FU-1&=&0,\lb{3.30}\\
D_1\phi\pm \ri D_2 \phi &=& 0,\lb{3.31} \\
D_1\psi\pm \ri D_2 \psi &=& 0.\lb{3.32}
\eea
Combining \eq{3.29} and \eq{3.30}, we get
\be\lb{ax6.15}
\tilde{F}_{12}=\mp\frac{b\frac{|\phi|^2-1}{|\phi|^2+1}+d\frac{|\psi|^2-1}{|\psi|^2+1}}{\sqrt{1-\frac{1}{b_1^2}\left(b\frac{|\phi|^2-1}{|\phi|^2+1}+d\frac{|\psi|^2-1}{|\psi|^2+1}\right)^2}}.
\ee
Consequently,  in view of \eq{4.1}, \eq{5.2}, and using the substitutions $u=\ln|\phi|^2$ and $v=\ln|\psi|^2$, we come up with the nonlinear elliptic  vortex equations
\bea
\Delta u&=&2a^2\frac{\e^u-1}{\e^u+1}+2ac\frac{\e^v-1}{\e^v+1}+\frac{2\left(b^2\frac{\e^u-1}{\e^u+1}+bd\frac{\e^v-1}{\e^v+1}\right)}{\sqrt{1-\frac{1}{b_1^2}\left(b\frac{\e^u-1}{\e^u+1}+d\frac{\e^v-1}{\e^v+1}\right)^2}}\nn\\
&&+4\pi \sum_{s=1}^{M_1}\delta_{q'_s}(x)-4\pi \sum_{s=1}^{N_1}\delta_{p'_s}(x),\lb{axx6.16}\\
\Delta v&=&2ac\frac{\e^u-1}{\e^u+1}+2c^2\frac{\e^v-1}{\e^v+1}+\frac{2\left(bd\frac{\e^u-1}{\e^u+1}+d^2\frac{\e^v-1}{\e^v+1}\right)}{\sqrt{1-\frac{1}{b_1^2}\left(b\frac{\e^u-1}{\e^u+1}+d\frac{\e^v-1}{\e^v+1}\right)^2}}\nn\\
&&+4\pi \sum_{s=1}^{M_2}\delta_{q''_s}(x)-4\pi \sum_{s=1}^{N_2}\delta_{p''_s}(x).\lb{axx6.17}
\eea
As in Section \ref{sec5}, integrating these equations enables us to arrive at the equations
\bea
a\int f_3\,\dd x+b\int\frac{f_4}{\sqrt{1-\frac{1}{b_1^2}f_4^2}}\,\dd x&=& -2\pi(M_1-N_1),
\\
c\int f_3\,\dd x+d\int\frac{f_4}{\sqrt{1-\frac{1}{b_1^2}f_4^2}}\,\dd x&=& -2\pi(M_2-N_2).
\eea
Solving these equations, we get from \eq{3.28} and \eq{ax6.15} the magnetic fluxes
\bea
\hat{\Phi}&=&\int \hat{F}_{12}\,\dd x=\mp\int f_3\,\dd x=\pm\frac{2\pi(d(M_1-N_1)-b(M_2-N_2))}{ad-bc},\\
\tilde{\Phi}&=&\int\tilde{F}_{12}\,\dd x=\mp\int\frac{f_4}{\sqrt{1-\frac1{b_1^2}f_4^2}}\,\dd x=\pm\frac{2\pi(a(M_2-N_2)-c(M_1-N_1))}{ad-bc},
\eea
which are identical to \eq{ax5.32} and \eq{ax5.33}. Thus, using \eq{ax5.34}, we arrive at \eq{ax5.35} again.

\section{Bogomol'nyi equations governing Maxwell vortices and Born--Infeld vortices and antivortices}\lb{sec7}
In Sections 3 and 4, we derived the Bogomol'nyi equations governing two species of vortices in product Born--Infeld and Maxwell--Born--Infeld theories. In Sections 5 and 6, we derived
the Bogomol'nyi equations accommodating
 {\em both} vortices and antivortices in  such theories.
In the next two sections, we derive the
 Bogomol'nyi equations governing  vortices {\em and} vortices--antivortices in product Maxwell--Born--Infeld theories as well.

In the product Maxwell--Born--Infeld theory, the static energy density reads
\be\lb{6.1}
\mathcal{H}=\frac{1}{2}\hat{F}_{12}^2+b_1^2\left(\sqrt{1+\frac{1}{b_1^2}\tilde{F}_{12}^2}-1\right)+\frac12|D_i\phi|^2
+\frac{2}{(1+|\psi|^2)^2}|D_i\psi|^2+V,
\ee
for which the Euler--Lagrange equations  are
\bea
D_i^2\phi&=&2\frac{\pa V}{\pa \overline{\phi}},\lb{6.2}\\
D_i\left(\frac{D_i\psi}{(|\psi|^2+1)^2}\right)&=& -\frac{2|D_i\psi|^2 \psi}{(|\psi|^2+1)^3}+\frac12\frac{\pa V}{\pa \overline{\psi}},\lb{6.3}\\
\pa_j\hat{F}_{ij}&=& \frac{\ri}{2}(a[\phi\overline{D_i\phi}-\overline{\phi}D_i\phi])+\frac{2 \ri(c[\psi\overline{D_i\psi}-\overline{\psi}D_i\psi])}{(1+|\psi|^2)^2},\lb{6.4}\\
\pa_j\left(\frac{ \tilde{F}_{ij}}{\sqrt{1+\frac{\tilde{F}_{12}^2}{b_1^2}}}\right)&=&\frac{\ri}{2}(b[\phi\overline{D_i\phi}-\overline{\phi}D_i\phi])+\frac{2 \ri(d[\psi\overline{D_i\psi}-\overline{\psi}D_i\psi])}{(1+|\psi|^2)^2}.
\lb{6.5}
\eea
On the other hand,  using \eq{x2.7} and \eq{ax2.11} with $\phi$ being replaced by $\psi$, we can rewrite the Hamiltonian density \eq{6.1} as
\bea
\mathcal{H}
&=&\frac{1}{2}\left(\hat{F}_{12}\pm\left(\frac{a}{2}(|\phi|^2-\xi)+c\frac{|\psi|^2-1}{|\psi|^2+1}\right)\right)^2\nn \\
&&+\frac{1}{2}\left(\tilde{F}_{12}\pm F\left(\frac{b}{2}(|\phi|^2-\xi)+d\frac{|\psi|^2-1}{|\psi|^2+1}\right)\right)^2 F^{-1}+\frac{b_1^2}{2}(FU-1)^2F^{-1}\nn \\
&&+\frac12|D_1 \phi \pm \ri D_2 \phi|^2+\frac{2|D_1 \psi \pm \ri D_2 \psi|^2 }{(1+|\psi|^2)^2}\nn \\
&&\pm \left(-\frac{|\phi|^2}{2}(a\hat{F}_{12}+b\tilde{F}_{12})\right)\pm \frac{\ri}{2} \left(\pa_1[\phi\overline{D_2 \phi}]-\pa_2[\phi\overline{D_1\phi}]\right)\pm \frac{|\phi|^2}{2}(a\hat{F}_{12}+b\tilde{F}_{12})\nn \\
&&\pm \left(\frac{-2|\psi|^2}{1+|\psi|^2}(c\hat{F}_{12}+d\tilde{F}_{12})\right)\pm \frac{2\ri(D_1\psi\overline{D_2 \psi}-\overline{D_1\psi}D_2 \psi)}{(1+|\psi|^2)^2}\nn \\
&&\pm\left(\frac{a\xi}{2}+c\right)\hat{F}_{12}\pm\left(\frac{b\xi}{2}+d\right)\tilde{F}_{12}\nn\\
&&-\frac{1}{2}\left(\frac{a}{2}(|\phi|^2-\xi)+c\frac{|\psi|^2-1}{|\psi|^2+1}\right)^2
-b_1^2(1-U)+V, \lb{6.6}
\eea
where $F$ and $U$ are
\be
F={\sqrt{1+\frac1{b_1^2}\tilde{F}_{12}^2}}, \quad U= \sqrt{1-\frac{1}{b_1^2}\left(\frac{b}{2}(|\phi|^2-\xi)+d\frac{|\psi|^2-1}{|\psi|^2+1}\right)^2}.
\ee
Hence we arrive at the potential density
\be\lb{6.7}
V=\frac{1}{2}\left(\frac{a}{2}(|\phi|^2-\xi)+c\frac{|\psi|^2-1}{|\psi|^2+1}\right)^2 +b_1^2\left(1-\sqrt{1-\frac{1}{b_1^2}\left(\frac{b}{2}(|\phi|^2-\xi)+d\frac{|\psi|^2-1}{|\psi|^2+1}\right)^2}\right).
\ee
It is interesting to note that this potential density mixes those of the Abelian Higgs theory potential and of harmonic map model potential in a
blended Higgs and  Born--Infeld model manner.  Using \eq{3.14} and \eq{6.7}  in \eq{6.6}, we arrive at
\bea
\mathcal{H}&=&\frac{1}{2}\left(\hat{F}_{12}\pm\left(\frac{a}{2}(|\phi|^2-\xi)+c\frac{|\psi|^2-1}{|\psi|^2+1}\right)\right)^2\nn \\
&&+\frac{1}{2}\left(\tilde{F}_{12}\pm F\left(\frac{b}{2}(|\phi|^2-\xi)+d\frac{|\psi|^2-1}{|\psi|^2+1}\right)\right)^2 F^{-1}+\frac{b_1^2}{2}(FU-1)^2F^{-1}\nn \\
&&+\frac12|D_1 \phi \pm \ri D_2 \phi|^2\pm \frac{\ri}{2} \left(\pa_1[\phi\overline{D_2 \phi}]-\pa_2[\phi\overline{D_1\phi}]\right)+\frac{2}{(1+|\psi|^2)^2}|D_1 \psi \pm \ri D_2 \psi|^2 \nn \\
&&\pm\left(\frac{a\xi}{2}+c\right)\hat{F}_{12}\pm\left(\frac{b\xi}{2}+d\right)\tilde{F}_{12}\pm J_{12}(\psi).\lb{6.8}
\eea
Thus the energy obtained enjoys the lower bound
\bea\lb{6.9}
E&=&\int_{\bfR^2}\mathcal{H}\,\dd x\nn\\
&\ge&\pm \left(\frac{a\xi}{2}+c\right)\int_{\bfR^2}\hat{F}_{12}\,\dd x\pm\left (\frac{b\xi}{2}+d\right)\int_{\bfR^2}\tilde{F}_{12}\,\dd x\pm \int_{\bfR^2} J_{12}(\psi)\,\dd x,\quad\quad
\eea
which is attained  if all the quadratic terms in \eq{6.8} vanish,  so that we derive the Bogomol'nyi equations:
\bea
\hat{F}_{12}\pm \left(\frac{a}{2}(|\phi|^2-\xi)+c\frac{|\psi|^2-1}{|\psi|^2+1}\right)&=&0,\lb{6.10}\\
\tilde{F}_{12}\pm F\left(\frac{b}{2}(|\phi|^2-\xi)+d\frac{|\psi|^2-1}{|\psi|^2+1}\right)&=&0,\lb{6.11}\\
FU-1&=&0,\lb{6.12}\\
D_1\phi\pm \ri D_2 \phi &=& 0, \lb {6.13}\\
D_1\psi\pm \ri D_2 \psi &=& 0. \lb{6.14}
\eea
Rewriting \eq{6.10} and combining \eq{6.11} and \eq{6.12}, we have the  vorticity field equations
\bea
\hat{F}_{12}&=&\mp\left(\frac{a}{2}(|\phi|^2-\xi)+c\frac{|\psi|^2-1}{|\psi|^2+1}\right),\lb{6.15a}\\
\tilde{F}_{12}&=&\mp \frac{\frac{b}{2}(|\phi|^2-\xi)+d\frac{|\psi|^2-1}{|\psi|^2+1}}{\sqrt{1-\frac{1}{b_1^2}\left(\frac{b}{2}(|\phi|^2-\xi)+d\frac{|\psi|^2-1}{|\psi|^2+1}\right)^2}}.\lb{6.15b}
\eea
As before, let the zeros of $\phi$  represent the vortices and the zeros and poles of $\psi$ the vortices and antivortices of the system, given by
\be
\phi: \left\{q'_1,\dots,q'_{M_1}\right\}; \quad \psi: \left\{{q}''_1,\dots,{q}''_{M_2}\right\},\quad
\left\{{p}''_1,\dots,{p}''_{N_2}\right\}.\lb{6.16}
\ee
Then, in view of \eq{4.1}, \eq{6.15a}, \eq{6.15b}, and the substitutions of the variables, $u=\ln|\phi|^2$ and $v=\ln|\psi|^2$, we see that the Bogomol'nyi equations \eq{6.10}--\eq{6.14}
are recast into the nonlinear elliptic equations:
\bea
\Delta u&=&a^2(\e^u-\xi)+2ac\frac{\e^v-1}{\e^v+1}+\frac{b^2(\e^u-\xi)+2bd\frac{\e^v-1}{\e^v+1}}{\sqrt{1-\frac{1}{b_1^2}\left(\frac{b}{2}(\e^u-\xi)+d\frac{\e^v-1}{\e^v+1}\right)^2}}\nn\\
&&+4\pi \sum_{s=1}^{M_1}\delta_{q'_s}(x),\lb{6.17}\\
\Delta v&=&ac(\e^u-\xi)+2c^2\frac{\e^v-1}{\e^v+1}+\frac{bd(\e^u-\xi)+2d^2\frac{\e^v-1}{\e^v+1}}{\sqrt{1-\frac{1}{b_1^2}\left(\frac{b}{2}(\e^u-\xi)+d\frac{\e^v-1}{\e^v+1}\right)^2}}\nn\\
&&+4\pi \sum_{s=1}^{M_2}\delta_{q''_s}(x)-4\pi \sum_{s=1}^{N_2}\delta_{p''_s}(x).\lb{6.18}
\eea
Integrating \eq{6.17} and \eq{6.18} yields the formulas
\bea
a\int f_5\,\dd x+b\int\frac{f_6}{\sqrt{1-\frac{1}{b_1^2}f_6^2}}\,\dd x&=& -2\pi M_1,\lb{x7.21}
\\
c\int f_5\,\dd x+d\int\frac{f_6}{\sqrt{1-\frac{1}{b_1^2}f_6^2}}\,\dd x&=& -2\pi(M_2-N_2),\lb{x7.22}
\eea
where $f_5=f_5(u,v)$ and $f_6=f_6(u,v)$ are defined by
\be\lb{ax7.23}
f_5=\frac{a}{2}(\e^u-\xi)+c\frac{\e^v-1}{\e^v+1},\quad f_6=\frac{b}{2}(\e^u-\xi)+d\frac{\e^v-1}{\e^v+1}.
\ee
By direct calculation, we derive from \eq{6.15a}, \eq{6.15b},  \eq{x7.21}, and \eq{x7.22} the magnetic fluxes
\bea
\hat{\Phi}&=&\int \hat{F}_{12}\,\dd x=\mp\int f_5\,\dd x=\pm\frac{2\pi(d M_1-b(M_2-N_2))}{ad-bc},\lb{6.19}\\
\tilde{\Phi}&=&\int\tilde{F}_{12}\,\dd x=\mp\int\frac{f_6}{\sqrt{1-\frac1{b_1^2}f_6^2}}\,\dd x=\pm\frac{2\pi(a(M_2-N_2)-c M_1)}{ad-bc}.\lb{6.20}
\eea

 Similarly, the quantity $J_{12}(\psi)$ also satisfies the second relation in \eq{ax5.34}.

Substituting this relation along with \eq{6.19} and \eq{6.20} into \eq{6.9}, we obtain the
quantized energy
\be\lb{7.26}
E=2\pi \left(\frac{\xi}{2} M_1+M_2+N_2\right),
\ee
in terms of numbers of the Maxwell vortices and Born--Infeld vortices and antivortices.

\section{Bogomol'nyi equations governing Born--Infeld vortices and Maxwell vortices and antivortices}\lb{sec8}

In Section \ref{sec7}, we derived the Bogomol'nyi equations governing the Maxwell vortices and Born--Infeld vortices and antivortices. To complete the study, here we
derive the Bogomol'nyi equations governing the Born--Infeld vortices and Maxwell vortices and antivortices. For this purpose,
we consider an alternative model whose static Hamiltonian energy density is
\be\lb{7.1}
\mathcal{H}=\frac{1}{2}\hat{F}_{12}^2+b_2^2\left(\sqrt{1+\frac{1}{b_2^2}\tilde{F}_{12}^2}-1\right)+\frac{2}{(1+|\phi|^2)^2}|D_i\phi|^2
+\frac12|D_i\psi|^2+V,
\ee
giving rise to the associated Euler--Lagrange equations
\bea
D_i\left(\frac{D_i\phi}{(|\phi|^2+1)^2}\right)&=& -\frac{2|D_i\phi|^2 \phi}{(|\phi|^2+1)^3}+\frac12\frac{\pa V(|\phi|^2,|\psi|^2)}{\pa \overline{\phi}}, \lb{7.2} \\
D_i^2\psi&=&2\frac{\pa V}{\pa \overline{\psi}}, \lb{7.3} \\
\pa_j\hat{F}_{ij}&=&\frac{2 \ri(a[\phi\overline{D_i\phi}-\overline{\phi}D_i\phi])}{(1+|\phi|^2)^2}+\frac{\ri}{2}(c[\psi\overline{D_i\psi}-\overline{\psi}D_i\psi]),
\lb{7.4}\\
\pa_j\left(\frac{ \tilde{F}_{ij}}{\sqrt{1+\frac{\tilde{F}_{12}^2}{b_2^2}}}\right)&=&\frac{2 \ri(b[\phi\overline{D_i\phi}-\overline{\phi}D_i\phi])}{(1+|\phi|^2)^2}+\frac{\ri}{2}(d[\psi\overline{D_i\psi}-\overline{\psi}D_i\psi]).
\lb{7.5}
\eea
On the other hand, using \eq{ax2.11} and \eq{x2.8}, we can rewrite \eq{7.1} as
\bea
\mathcal{H}
&=&\frac{1}{2}\left(\hat{F}_{12}\pm\left(a\frac{|\phi|^2-1}{|\phi|^2+1}+\frac{c}{2}(|\psi|^2-\zeta)\right)\right)^2\nn \\
&&+\frac{1}{2}\left(\tilde{F}_{12}\pm F\left(b\frac{|\phi|^2-1}{|\phi|^2+1}+\frac{d}{2}(|\psi|^2-\zeta)\right)\right)^2 F^{-1}+\frac{b_2^2}{2}(FU-1)^2F^{-1}\nn \\
&&+\frac{2}{(1+|\phi|^2)^2}|D_1 \phi \pm \ri D_2 \phi|^2+\frac12|D_1 \psi \pm \ri D_2 \psi|^2\nn \\
&&\pm \left(\frac{-2|\phi|^2}{1+|\phi|^2}(a\hat{F}_{12}+b\tilde{F}_{12})\right)\pm \frac{2\ri}{(1+|\phi|^2)^2}(D_1\phi\overline{D_2 \phi}-\overline{D_1\phi}D_2 \phi) \nn \\
&&\pm \left(-\frac{|\psi|^2}{2}(c\hat{F}_{12}+d\tilde{F}_{12})\right)\pm \frac{\ri}{2} \left(\pa_1[\psi\overline{D_2 \psi}]-\pa_2[\psi\overline{D_1\psi}]\right)\pm \frac{|\psi|^2}{2}(c\hat{F}_{12}+d\tilde{F}_{12})\nn \\
&&\pm\left(a+\frac{c}{2}\zeta\right)\hat{F}_{12}\pm\left(b+\frac{d}{2}\zeta\right)\tilde{F}_{12}\nn\\
&&-\frac{1}{2}\left(a\frac{|\phi|^2-1}{|\phi|^2+1}+\frac{c}{2}(|\psi|^2-\zeta)\right)^2
-b_2^2(1-U)+V, \lb{7.6}
\eea
where $F$ and $U$ are
\be
F={\sqrt{1+\frac1{b_2^2}\tilde{F}_{12}^2}}, \quad U= \sqrt{1-\frac{1}{b_2^2}\left(b\frac{|\phi|^2-1}{|\phi|^2+1}+\frac{d}{2}(|\psi|^2-\zeta)\right)^2}.
\ee
Thus, if we define the potential density $V=V(|\phi|^2,|\psi|^2)$ to be
\be\lb{7.7}
V=\frac{1}{2}\left(a\frac{|\phi|^2-1}{|\phi|^2+1}+\frac{c}{2}(|\psi|^2-\zeta)\right)^2 +b_2^2\left(1-\sqrt{1-\frac{1}{b_2^2}\left(b\frac{|\phi|^2-1}{|\phi|^2+1}+\frac{d}{2}(|\psi|^2-\zeta)\right)^2}\right),
\ee
then inserting \eq{3.13} and \eq{7.7}  into \eq{7.6} leads us to get the familiar Bogomol'nyi structure
\bea
\mathcal{H}&=&\frac{1}{2}\left(\hat{F}_{12}\pm\left(a\frac{|\phi|^2-1}{|\phi|^2+1}+\frac{c}{2}(|\psi|^2-\zeta)\right)\right)^2\nn \\
&&+\frac{1}{2}\left(\tilde{F}_{12}\pm F\left(b\frac{|\phi|^2-1}{|\phi|^2+1}+\frac{d}{2}(|\psi|^2-\zeta)\right)\right)^2 F^{-1}+\frac{b_2^2}{2}(FU-1)^2F^{-1}\nn \\
&&+\frac{2}{(1+|\phi|^2)^2}|D_1 \phi \pm \ri D_2 \phi|^2\pm \frac{\ri}{2} \left(\pa_1[\psi\overline{D_2 \psi}]-\pa_2[\psi\overline{D_1\psi}]\right)+\frac12|D_1 \psi \pm \ri D_2 \psi|^2 \nn \\
&&\pm\left(a+\frac{c}{2}\zeta\right)\hat{F}_{12}\pm\left(b+\frac{d}{2}\zeta\right)\tilde{F}_{12}\pm J_{12}(\phi).\lb{7.8}
\eea
Thus the energy obtained enjoys the lower bound
\bea\lb{7.9}
E&=&\int_{\bfR^2}\mathcal{H}\,\dd x\nn\\
&\ge&\pm\left (a+\frac{c}{2}\zeta\right)\int_{\bfR^2}\hat{F}_{12}\,\dd x\pm \left(b+\frac{d}{2}\zeta\right)\int_{\bfR^2}\tilde{F}_{12}\,\dd x\pm \int_{\bfR^2} J_{12}(\phi)\,\dd x,\quad\quad
\eea
which is saturated if all the quadratic terms in \eq{7.8} vanish, rendering us the Bogomol'nyi equations
\bea
\hat{F}_{12}\pm \left(a\frac{|\phi|^2-1}{|\phi|^2+1}+\frac{c}{2}(|\psi|^2-\zeta)\right)&=&0,\lb{7.10}\\
\tilde{F}_{12}\pm F\left(b\frac{|\phi|^2-1}{|\phi|^2+1}+\frac{d}{2}(|\psi|^2-\zeta)\right)&=&0,\lb{7.11}\\
FU-1&=&0,\lb{7.12}\\
D_1\phi\pm \ri D_2 \phi &=& 0, \lb {7.13}\\
D_1\psi\pm \ri D_2 \psi &=& 0. \lb{7.14}
\eea
With \eq{7.10} and combining \eq{7.11} and \eq{7.12}, we have the following vorticity field equations
\bea
\hat{F}_{12}&=&\mp \left(a\frac{|\phi|^2-1}{|\phi|^2+1}+\frac{c}{2}(|\psi|^2-\zeta)\right),\lb{7.15a}\\
\tilde{F}_{12}&=&\mp \frac{b\frac{|\phi|^2-1}{|\phi|^2+1}+\frac{d}{2}(|\psi|^2-\zeta)}{\sqrt{1-\frac{1}{b_2^2}\left(b\frac{|\phi|^2-1}{|\phi|^2+1}+\frac{d}{2}(|\psi|^2-\zeta)\right)^2}}.\lb{7.15b}
\eea
Therefore, by assigning the zeros and poles to $\phi$ and zeros to $\psi$,
\be
\phi: \left\{q'_1,\dots,q'_{M_1}\right\},\quad \left\{{p}'_1,\dots,{p}'_{N_1}\right\}; \quad \psi: \left\{{q}''_1,\dots,{q}''_{M_2}\right\},
\lb{7.16}
\ee
to represent vortices and antivortices generated by $\phi$ and vortices generated by $\psi$, respectively,
and applying the relations in \eq{4.1},  the substitutions $u=\ln|\phi|^2$ and $v=\ln|\psi|^2$ enable us to arrive at the nonlinear elliptic equations
\bea
\Delta u&=&2a^2\frac{\e^u-1}{\e^u+1}+ac(\e^v-\zeta)+\frac{2b^2\frac{\e^u-1}{\e^u+1}+bd(\e^v-\zeta)}{\sqrt{1-\frac{1}{b_2^2}\left(b\frac{\e^u-1}{\e^u+1}+\frac{d}{2}(\e^v-\zeta)\right)^2}}\nn\\
&&+4\pi \sum_{s=1}^{M_1}\delta_{q'_s}(x)-4\pi \sum_{s=1}^{N_1}\delta_{p'_s}(x),\lb{7.17}\\
\Delta v&=&2ac\frac{\e^u-1}{\e^u+1}+c^2(\e^v-\zeta)+\frac{2bd\frac{\e^u-1}{\e^u+1}+d^2(\e^v-\zeta)}{\sqrt{1-\frac{1}{b_2^2}\left(b\frac{\e^u-1}{\e^u+1}+\frac{d}{2}(\e^v-\zeta)\right)^2}}\nn\\
&&+4\pi \sum_{s=1}^{M_2}\delta_{q''_s}(x).\lb{7.18}
\eea
Integrating \eq{7.17} and \eq{7.18} gives us the expressions
\bea
a\int f_7\,\dd x+b\int\frac{f_8}{\sqrt{1-\frac{1}{b_2^2}f_8^2}}\,\dd x&=& -2\pi (M_1-N_1),\lb{ax7.19}
\\
c\int f_7\,\dd x+d\int\frac{f_8}{\sqrt{1-\frac{1}{b_2^2}f_8^2}}\,\dd x&=& -2\pi M_2,\lb{ax7.20}
\eea
where $f_7=f_7(u,v)$ and $f_8=f_8(u,v)$ are given by
\be\lb{ax7.18}
f_7=a\frac{\e^u-1}{\e^u+1}+\frac{c}{2}(\e^v-\zeta),\quad f_8=b\frac{\e^u-1}{\e^u+1}+\frac{d}{2}(\e^v-\zeta).
\ee
In view of \eq{7.15a} and \eq{7.15b}, we obtain the magnetic fluxes
\bea
\hat{\Phi}&=&\int \hat{F}_{12}\,\dd x=\mp\int f_7\,\dd x=\pm\frac{2\pi(d (M_1-N_1)-b M_2)}{ad-bc},\lb{7.19}\\
\tilde{\Phi}&=&\int\tilde{F}_{12}\,\dd x=\mp\int\frac{f_8}{\sqrt{1-\frac1{b_2^2}f_8^2}}\,\dd x=\pm\frac{2\pi(a M_2-c (M_1-N_1))}{ad-bc}.\lb{7.20}
\eea
Inserting the first relation in \eq{ax5.34}, \eq{7.19}, and \eq{7.20} into \eq{7.9}, we get the quantized energy
\be
E=2\pi \left(M_1+N_1+\frac{\zeta}{2} M_2\right),
\ee
expressed in terms of the numbers of the Born--Infeld vortices and Maxwell vortices and antivortices, in contrast to the result \eq{7.26}.

Note that, strictly speaking, the vortices and antivortices generated in the coupled Maxwell--Born--Infeld theories, \eq{6.1} and \eq{7.1}, are mixed, as shown in the pairs of
formulas, \eq{6.19}, \eq{6.20}, and \eq{7.19}, \eq{7.20}, rather than carrying a separate individual Maxwell or Born--Infeld theory signature. They are justified to be referred to as
the  Maxwell or Born--Infeld
vortices only when their Born--Infeld or Maxwell model partner is switched off or trivialized such that the theory becomes a model with a single gauge field and a single scalar field.

\section{Thermodynamics of pinned vortices}\lb{sec9}

Having established the Bogomol'nyi structure and exact energy spectra of the vortex systems, we turn to their statistical-mechanical properties. In this section we analyze the thermodynamics of pinned vortices and vortex-antivortex pairs in the full-plane setting. The linear dependence of the energy on topological quantum numbers allows the canonical partition functions to be evaluated explicitly. We derive closed-form expressions for the internal energy, heat capacity, and magnetization, and we examine their behavior in the low- and high-temperature regimes. These results provide a transparent illustration of how topology and self-duality shape the macroscopic thermal response of nonlinear gauge systems.

For both conceptual clarity and practical relevance, it is often advantageous to study the thermodynamics of vortices under the simplifying assumption that their positions are fixed.  In many physical systems---such as type-II superconductors with strong pinning, optical or atomic lattices, or engineered superconducting arrays---the vortices are immobilized by an underlying lattice or by impurities, forming a solid-like structure in which positional degrees of freedom play little dynamical role.  By treating the vortex cores as fixed and focusing on their internal or topological characteristics (vortex versus antivortex), one captures the essential magnetic and thermodynamic behavior while avoiding the complications of collective motion and long-range interactions.  This pinned-lattice model thus provides a clean theoretical setting for analyzing equilibrium properties, magnetization, and response to external fields, while remaining closely aligned with experimentally realizable situations.

In a two-species system governed by a $U(1)\times U(1)$ gauge structure, each vortex species carries its own conserved flux and order-parameter phase, yet the two are coupled through electromagnetic or Born--Infeld type nonlinearities.  Studying the thermodynamics of such a system is motivated both by physical realism and by theoretical richness.  Physically, many multicomponent condensates---such as multiband superconductors, two-component Bose--Einstein condensates, and certain cosmic-string or brane models---exhibit precisely this dual $U(1)$ structure, where interspecies coupling gives rise to composite vortex, antivortex, and dyonic configurations.  Thermodynamically, the coexistence of two flux species introduces new variables (two magnetic fields, two chemical potentials, and intercomponent correlations), making it possible to explore cross-susceptibilities, mutual screening, and mixed-phase transitions within a unified statistical framework.  Moreover, analyzing the equilibrium behavior of a pinned two-species vortex lattice provides a tractable and illuminating model for understanding how nonlinear field theories with multiple gauge sectors store and redistribute energy and flux under external perturbations, offering insight into both condensed-matter and high-energy analogs.

In this section, we study the thermodynamics of the systems of vortices in the two-species Maxwell and Born--Infeld model \eq{ax4.1} and vortices and antivortices in the two-species
Born--Infeld model \eq{3.10} as illustrative examples. For simplicity, we assume zero chemical potential and a constant applied magnetic field, $B\geq0$.

\subsection{The model \eq{ax4.1}}

In the presence of $B$, the Hamiltonian energy density \eq{ax4.1} is updated into the form
\be\lb{11.1}
{\cal H}_B
=\frac{1}{2}\hat{F}_{12}^2+b_1^2\left(\sqrt{1+\frac{1}{b_1^2}\tilde{F}_{12}^2}-1\right)+\frac12|D_i\phi|^2+\frac12|D_i\psi|^2 +V-\hat{F}_{12} B
-\tilde{F}_{12}B,
\ee
where we assume paramagnetism and the potential density function $V$ is as defined by \eq{ax4.8} to achieve
a Bogomol'nyi structure. Hence, for the two-species system of vortices given by the data \eq{4.0}, we can use \eq{4.10c}, \eq{ax4.22}, and \eq{ax4.23} and observe self-duality for
convenience to get
the energy
\bea\lb{11.2}
E_{B,M_1,M_2}&=&\pi(\xi M_1+\zeta M_2)-\frac{2\pi B}{ad-bc}((dM_1-bM_2)+(aM_2-cM_1))\nn\\
&=&\pi M_1\left(\xi-\frac{2(d-c)B}{ad-bc}\right)+\pi M_2\left(\zeta-\frac{2(a-b)B}{ad-bc}\right).
\eea
To proceed further, we subsequently consider a weak applied field situation such that
\be
\sigma_1\equiv\xi-\frac{2(d-c)B}{ad-bc}>0,\quad \sigma_2\equiv\zeta-\frac{2(a-b)B}{ad-bc}>0.
\ee
Therefore we see from \eq{11.2} that, at zero-temperature, the least-energy principle imposes the condition
\be
M_1=M_2=0.
\ee
In other words, no vortices are present and magnetic screening takes place. This phenomenon is known as the Meissner effect, of course.
At a finite temperature, however, thermodynamic fluctuations enable vortices to appear as we now study.

With the energy spectrum given in \eq{11.2}, we can express the partition function at the absolute temperature $T>0$ as
\bea
Z&=&\sum_{M_1,M_2\geq0} \e^{-\beta E_{B,M_1,M_2}}=\left(\sum_{M_1\geq0} \e^{-\beta \pi \sigma_1 M_1}\right)\left(\sum_{M_2\geq0} \e^{-\beta \pi \sigma_2 M_2}\right)\nn\\
&=&\frac1{(1-\e^{-\beta\pi\sigma_1})(1-\e^{-\beta\pi\sigma_2})}\equiv Z_1 Z_2,\quad\beta=\frac1{\kb T},
\eea
where $\kb$ is the Boltzmann constant. Hence we get the internal energy of the system
\bea\lb{ax9.6}
U&=&\sum_{M_1,M_2\geq0}E_{B,M_1,M_2} \frac{ \e^{-\beta E_{B,M_1,M_2}}}Z=-\frac{\pa \ln Z}{\pa\beta}\nn\\
&=&\frac{\pi\sigma_1}{\e^{\beta\pi\sigma_1}-1}+\frac{\pi\sigma_2}{\e^{\beta\pi\sigma_2}-1},
\eea
which gives rise to the asymptotic properties
\be
U\to 0,\quad T\to 0;\quad U\to \infty,\quad T\to\infty,
\ee
 of the two-species vortex system,
and the associated heat capacity
\be\lb{11.7}
C_{\rm{V}}=\frac{\pa U}{\pa T}=\frac{\pi^2}{\kb T^2}\left(\frac{\sigma_1^2\e^{\beta\pi\sigma_1}}{(\e^{\beta\pi\sigma_1}-1)^2}+\frac{\sigma_2^2\e^{\beta\pi\sigma_2}}{(\e^{\beta\pi\sigma_2}-1)^2}\right),
\ee
which enjoys the asymptotic properties
\be\lb{abx11.7}
C_{\rm{V}}\to0,\quad T\to0;\quad C_{\rm{V}}\to 2\kb,\quad T\to\infty,
\ee
independent of the applied field $B$.
Furthermore, with the joint magnetic flux
\bea\lb{ax9.10}
\Phi_{M_1,M_2}&=&\hat{\Phi}+\tilde{\Phi}=\frac{2\pi}{ad-bc}((dM_1-bM_2)+(aM_2-cM_1))\nn\\
&=&\frac{2\pi}{ad-bc}((d-c)M_1+(a-b)M_2)
\eea
at the vortex numbers $M_1$ and $M_2$,
we may calculate the magnetization
\bea\lb{11.10}
M(T)&=&\sum_{M_1,M_2\geq0} \Phi_{M_1,M_2}\frac{ \e^{-\beta E_{B,M_1,M_2}}}Z\nn\\
&=&\frac{2\pi}{ad-bc}\left(\sum_{M_1,M_2\geq0}( (d-c)M_1+(a-b)M_2)\frac{ \e^{-\beta \pi(\sigma_1M_1+\sigma_2M_2)}}Z\right)\nn\\
&=&-\frac{2\pi}{ad-bc}\left(\frac{(d-c)}{\pi\sigma_1}\frac{\pa\ln Z_1}{\pa\beta}+\frac{(a-b)}{\pi\sigma_2}\frac{\pa\ln Z_2}{\pa\beta}\right)\nn\\
&=&\frac{2\pi}{ad-bc}\left(\frac{d-c}{\e^{\beta\pi\sigma_1}-1}+\frac{a-b}{\e^{\beta\pi\sigma_2}-1}\right).
\eea
This expression leads us to the following scenarios:

\begin{enumerate}

\item[(i)] We have $M(T)\to0$ as $T\to0$. That is, there is no spontaneous magnetization at zero temperature, a phenomenon consistent with the Meissner effect.

\item[(ii)] We have
\bea\lb{ax9.12}
&&M(T)
=\frac{2k_B T}{ad-bc}
\left(
\frac{d-c}{\xi-\tfrac{2B(d-c)}{ad-bc}}
+\frac{a-b}{\zeta-\tfrac{2B(a-b)}{ad-bc}}
\right)
-\frac{\pi}{ad-bc}\,(a+d-b-c)\nn\\
&&+\frac{\pi^2\left[
(d-c)\xi+(a-b)\zeta
-\frac{2B}{ad-bc}\big((d-c)^2+(a-b)^2\big)
\right]}{6k_B T\,(ad-bc)}
+\mbox{O}(T^{-3}),
\eea
for $T\gg1$. In other words, $M(T)\sim \kb T$ up to a constant factor in high-$T$ limit. This model-specific growth law indicates that, in this pinned, unlimited-occupancy Boltzmann sum
ensemble, the number of vortex excitations is unbounded and becomes thermally populated as the energy cost for vortex creation is washed out.

\item[(iii)] If $B=0$, we have $\sigma_1=\xi$ and $\sigma_2=\zeta$. Thus \eq{11.10} yields the result
\be\lb{ax9.13}
M(T)=\frac{2\pi}{ad-bc}\left(\frac{d-c}{\e^{\beta\pi\xi}-1}+\frac{a-b}{\e^{\beta\pi\zeta}-1}\right).
\ee
Hence  spontaneous magnetization occurs in the model if and only if
\be
\frac{c-d}{\e^{\beta\pi\xi}-1}\neq \frac{a-b}{\e^{\beta\pi\zeta}-1}.
\ee
For example, if $\xi=\zeta$, then spontaneous magnetization happens at any temperature if and only if $a-b\neq c-d$.
\end{enumerate}

\subsection{The model \eq{3.10}}

We next consider the model \eq{3.10} such that
\bea
{\cal H}_B&=&b_1^2\left(\sqrt{1+\frac{1}{b_1^2}\hat{F}_{12}^2}-1\right)+b_2^2\left(\sqrt{1+\frac{1}{b_2^2}\tilde{F}_{12}^2}-1\right)\nn\\
&&+\frac{2|D_i\phi|^2 }{(1+|\phi|^2)^2}
+\frac{2|D_i\psi|^2}{(1+|\psi|^2)^2}+V-\hat{F}_{12} B
-\tilde{F}_{12}B,
\eea
 where the potential density function $V$ is given by \eq{ax5.10}. Thus the energy of the system of the two species of vortices and antivortices, with
the choice of self-duality for convenience, described by the data sets stated in
\eq{5.2} is
\be\lb{ax9.16}
E_{B,M_1,M_2,N_1,N_2}=2\pi(\sigma^-_1 M_1+\sigma^+_1 N_1+\sigma^-_2 M_2+\sigma^+_2 N_2),
\ee
where
\be\lb{ax9.17}
\sigma^\pm_1=1\pm\frac{(d-c)B}{ad-bc},\quad \sigma^\pm_2=1\pm\frac{(a-b)B}{ad-bc},
\ee
are assumed to be positive quantities subject to the weak applied field condition. As a consequence, we obtain the associated partition function
\bea
Z&=&\sum_{M_1,N_1,M_2,N_2\geq0} \e^{-\beta E_{B,M_1,M_2,N_1,N_2}}\nn\\
&=&\frac1{(1-\e^{-\beta 2\pi\sigma_1^-})}\frac1{(1-\e^{-\beta 2\pi\sigma_1^+})}\frac1{(1-\e^{-\beta 2\pi\sigma_2^-})}\frac1{(1-\e^{-\beta 2\pi\sigma_2^+})}\nn\\
&\equiv&
Z^-_1Z^+_1Z^-_2Z^+_2.
\eea
Thus we see that the internal energy of the system is
\bea
U&=&\sum_{M_1,N_1,M_2,N_2\geq0}  E_{B,M_1,M_2,N_1,N_2}\frac{\e^{-\beta E_{B,M_1,M_2,N_1,N_2}}}Z=-\frac{\pa\ln Z}{\pa\beta}\nn\\
&=&2\pi\left(\frac{\sigma^-_1}{\e^{\beta2\pi\sigma^-_1}-1}+\frac{\sigma^+_1}{\e^{\beta2\pi\sigma^+_1}-1}+\frac{\sigma^-_2}{\e^{\beta2\pi\sigma^-_2}-1}+\frac{\sigma^+_2}{\e^{\beta2\pi\sigma^+_2}-1}\right).
\eea
Hence we obtain the heat capacity of the system:
\bea
&&C_{\rm{V}}=\frac{\pa U}{\pa T}\nn\\
&&=\frac{(2\pi)^2}{\kb T^2}\left(\frac{(\sigma^-_1)^2\e^{\beta2\pi\sigma_1^-}}{(\e^{\beta2\pi\sigma_1^-}-1)^2}
+\frac{(\sigma^+_1)^2\e^{\beta2\pi\sigma_1^+}}{(\e^{\beta2\pi\sigma_1^+}-1)^2}+\frac{(\sigma^-_2)^2\e^{\beta2\pi\sigma_2^-}}{(\e^{\beta2\pi\sigma_2^-}-1)^2}
+\frac{(\sigma^+_2)^2\e^{\beta2\pi\sigma_2^+}}{(\e^{\beta2\pi\sigma_2^+}-1)^2}\right),\quad\quad\quad
\eea
such that there hold the asymptotic results
\be
C_{\rm{V}}\to0,\quad T\to0;\quad C_{\rm{V}}\to 4\kb,\quad T\to\infty,
\ee
similar to \eq{abx11.7}. Moreover, since for the vortex numbers $M_1,M_2$ and antivortex numbers $N_1,N_2$ the joint magnetic flux following from \eq{ax5.32} and \eq{ax5.33} is
\be
\Phi_{M_1,M_2,N_1,N_2}=\frac{2\pi}{ad-bc}\left((d-c)(M_1-N_1)+(a-b)(M_2-N_2)\right),
\ee
we get the associated magnetization
\bea\lb{11.22}
M(T)&=&\sum_{M_1,M_2,N_1,N_2\geq0}\Phi_{M_1,M_2,N_1,N_2}\frac{\e^{-\beta E_{B,M_1,M_2,N_1,N_2}}}Z=
\frac1\beta\frac{\pa \ln Z}{\pa B}\nn\\
&=&\frac{2\pi}{ad-bc}\left(\left[\frac{d-c}{\e^{\beta2\pi\sigma_1^-}-1}-\frac{d-c}{\e^{\beta2\pi\sigma_1^+}-1}\right]+
\left[\frac{a-b}{\e^{\beta2\pi\sigma_2^-}-1}-\frac{a-b}{\e^{\beta2\pi\sigma_2^+}-1}\right]\right).\quad\quad\quad
\eea
As before, we have $M(T)\to0$ as $T\to0$, indicating again the presence of an analogous Meissner effect at zero temperature due to a thermodynamic suppression of the magnetic flux
realized by pinned vortices and antivortices. Besides, we get from \eq{11.22} the expansion
\bea\lb{ax9.24}
M(T)&=&\frac{2\kb T B}{(ad-bc)^2}\left(\frac{(d-c)^2}{1-\left(\frac{(d-c)B}{ad-bc}\right)^2}+\frac{(a-b)^2}{1-\left(\frac{(a-b)B}{ad-bc}\right)^2}\right)\nn\\
&&-\frac{(2\pi)^2 B}{6(ad-bc)^2 \kb T}((d-c)^2+(a-b)^2)+\mbox{O}(T^{-3}),\quad T\gg1.
\eea
These behaviors are similar to those of the system of two-species vortices described by the model \eq{11.1}.

However, setting $B=0$ in the expression \eq{11.22}, we have $M(T)=0$, indicating the absence of spontaneous magnetization for any combinations of the charge parameters
$a,b,c,d$. This phenomenon is in sharp contrast to what happens in the two-species vortex system.

\medskip

It is worth emphasizing an instructive distinction between the high-temperature behaviors exhibited in \eq{ax9.12} and \eq{ax9.24}. In both cases, the magnetization $M(T)$ grows linearly with temperature as $T\to\infty$, a feature that reflects the unbounded excitation of topological degrees of freedom in the canonical ensemble. However, the underlying physical mechanisms responsible for this growth are fundamentally different. In the vortex-only setting leading to \eq{ax9.12}, the magnetization persists even in the absence of an applied field $B$. This behavior may be interpreted as a predetermined or biased magnetization inherent to the model itself, arising from the fact that all admissible configurations carry vortices of the same orientation. The absence of antivortices breaks the symmetry between positive and negative flux sectors at the outset, so that thermal excitation amplifies an intrinsic imbalance rather than responding to an external bias.

By contrast, the vortex-antivortex model underlying \eq{ax9.24} is symmetric with respect to the sign of the magnetic flux when $B=0$, and this symmetry forces the magnetization to vanish identically in the absence of an applied field. In this case, the linear growth of $M(T)$ at high temperature is genuinely field-driven: the applied magnetic field $B$ breaks the vortex-antivortex symmetry and selects a preferred orientation, and thermal excitation then enhances this externally induced bias. Consequently, the magnetization depends linearly on both $T$ and $B$, and disappears when $B=0$. From this perspective, the distinction between \eq{ax9.12} and \eq{ax9.24} is natural and physically transparent: the former reflects an intrinsic topological bias built into the configuration space, while the latter represents a conventional paramagnetic response in which magnetization emerges only through symmetry breaking by the applied field.

As we shall see in Section \ref{sec10}, the introduction of compact periodic geometry and the associated Bradlow type bounds further refine and enrich this distinction
by modifying the high-temperature thermodynamic response.

\section{Vortex condensation thermodynamics}\lb{sec10}

This section extends the thermodynamic analysis to a compact doubly periodic domain, which serves as a prototype for vortex lattices and the  Abrikosov type condensates. In this setting, geometric constraints give rise to the Bradlow type bounds that restrict the admissible vortex numbers and modify the thermodynamic behavior. We show how these bounds are reflected in the partition functions and lead to qualitative differences compared with the full-plane case, including altered high-temperature limits and possible suppression of thermal responses. The results underscore the interplay between geometry, topology, and thermodynamics in multi-component Born--Infeld vortex systems.

Specifically, so far, we have considered pinned vortices in a full plane setting. However,
in studying vortex condensation in superfluids or superconductors, it is essential to work over a bounded region modeling a finite fundamental cell domain with periodic boundary conditions in both spatial directions in the sense of 't Hooft \cite{Ybook,'tH} which realizes the lattice structure of a mixed state conceptualized by Abrikosov \cite{Ab}. Mathematically, the systems of
the vortex equations are now defined over a closed surface represented as a flat torus \cite{Ybook,WY}. Such a structure introduces some new features to the theories. For example, the size
of the flat torus or doubly periodic domain could  impose some upper bounds to the vortex numbers \cite{WY,N,B}, sometimes referred to as the Bradlow bound \cite{Manton,Gud}.
For a system of vortices and antivortices, on the other hand, such a bound is imposed on the difference of the number of vortices and the number of antivortices
 \cite{SSY,XY}. In this section, we explore the implication of these bounds to vortex thermodynamics. For comparison, we shall focus on the models \eq{ax4.1} and \eq{3.10} as examples.
Naturally and technically, the ranges of the charge parameters $a,b,c,d$ will play some subtle roles in this setting of discussion. For convenience and definiteness, we shall assume the positivity condition
\be\lb{c11.23}
a,b,c,d>0.
\ee
In view of \eq{ax3.1} and \eq{a2.6}, it is clear that the condition \eq{c11.23} also covers all the charge parameter cases
\be
(a,c,b,d)\sim (--++),(++--),(----),
\ee
through the sign flipping of either $\hat{A}_i\mapsto -\hat{A}_i$ or $\tilde{A}_i\mapsto -\tilde{A}_i$, or both.

\subsection{The vortex model \eq{ax4.1}}

For the vortex model \eq{ax4.1} defined
over a doubly-periodic domain $\Omega$, we may use the background functions $u'_0, u''_0$ defined by \eq{a4.4} and the substitutions $u=u'_0+w'$ and $v=u''_0+w''$ to recast \eq{4.11} and \eq{4.12} into the equations
\bea
\Delta w'&=&\left(a^2+\frac{b^2}{\sqrt{1-\frac{1}{4b_1^2}{f}_2^2}}\right)(\mathrm{e}^{u'_{0}+w'}-\xi)\nn \\
&&+\left(ac+\frac{bd}{\sqrt{1-\frac{1}{4b_1^2}{f}_2^2}}\right)(\mathrm{e}^{u''_{0}+w''}-\zeta)+\frac{4 \pi M_1}{|\Om|},\lb{e1}\\
\Delta w''&=&\left(ac+\frac{bd}{\sqrt{1-\frac{1}{4b_1^2}{f}_2^2}}\right)(\mathrm{e}^{u'_{0}+w'}-\xi)\nn \\
&&+\left(c^2+\frac{d^2}{\sqrt{1-\frac{1}{4b_1^2}{f}_2^2}}\right)(\mathrm{e}^{u''_{0}+w''}-\zeta)+\frac{4\pi M_2}{|\Om|},\lb{e2}
\eea
where $f_2$ is defined by \eq{e4}.
Integrating \eq{e1} and \eq{e2} over $\Om$ yields equations analogous to \eq{ax4.20} and \eq{ax4.21}, with $u$ and $v$ replaced by $u'_0+w'$ and $u''_0+w''$, respectively. Thus \eq{ax4.20} becomes
\be\lb{a11.22}
\frac{4\pi(b M_2-d M_1)}{ad-bc}+(a\xi+c\zeta)|\Om|=\int_{\Om}\left(a\e^{u'_0+w'}+c\e^{u''_0+w''}\right)\dd x,
\ee
which leads to the obvious necessary condition
\be\lb{b11.23}
\frac{4\pi(b M_2-d M_1)}{ad-bc}+(a\xi+c\zeta)|\Om|>0.
\ee
On the other hand, with $t_1=\e^{u'_0+w'}$ and $t_2=\e^{u''_0+w''}$, we may rewrite the integrand on the left-hand side of \eq{ax4.21} as
\be
g(t_1,t_2)=\frac{\left(b(t_1-\xi)+d(t_2-\zeta)\right)}{\sqrt{1-\frac{1}{4b_1^2}\left(b(t_1-\xi)+d(t_2-\zeta)\right)^2 }},
\ee
which has positive partial derivatives
\be\lb{11.23}
\left(\frac{\pa}{\pa t_1},\frac{\pa}{\pa t_2}\right)g=\frac{1}{(1-\frac{1}{4b_1^2}\left(b(t_1-\xi)+d(t_2-\zeta)\right)^2)^{\frac32}}(b,d).
\ee
In view of \eq{ax4.21} over $\Om$ and  \eq{11.23},  we have
\bea\lb{a11.23}
\frac{4\pi (cM_1-aM_2)}{ad-bc}&=&\int_{\Om}g(t_1,t_2)\,\dd x>\int_{\Om}g(0,0)\, \dd x\nn\\
&=&-\frac{|\Om|(b\xi+d\zeta)}{\sqrt{1-\frac1{4b_1^2}(b\xi+d\zeta)^2}}.
\eea
Combining \eq{b11.23} and \eq{a11.23},  we obtain the bounds
\bea
M_1&<&\frac{|\Om|}{4\pi}\left((a^2\xi+ac\zeta)+\frac{b^2\xi+bd\zeta}{\sqrt{1-\frac{1}{4b_1^2}(b\xi+d\zeta)^2}}\right),\lb{a11.25}\\
M_2&<&\frac{|\Om|}{4\pi}\left((ac\xi+c^2\zeta)+\frac{bd\xi+d^2\zeta}{\sqrt{1-\frac{1}{4b_1^2}(b\xi+d\zeta)^2}}\right),\lb{b11.25}
\eea
which are the upper bounds of the Bradlow type on the vortex numbers of the two species of vortices of the model \eq{ax4.1}. Since these bounds are strict inequalities for the
existence of solutions, we must consider the subtleties they impose. For this purpose, we denote by $K_1$ and $K_2$ the right-hand sides of \eq{a11.25} and \eq{b11.25}, respectively, and introduce
the notation
\be\lb{ax10.12}
M_i^0=\mbox{integer part of $K_i$ or $K_i-1$ if $K_i$ is already an integer},\quad i=1,2.
\ee
This notation recasts \eq{a11.25} and \eq{b11.25} into the condition
\be\lb{ax10.13}
M_i\le M_i^0,\quad i=1,2.
\ee

Now, in the presence of a constant magnetic field $B$, the energy formula
\eq{11.2} still holds.  Thus, with \eq{ax10.13},  we obtain the partition function
\bea\lb{11.26}
Z&=&\sum_{M_1,M_2\geq0}^{M_1\le M^0_1, M_2\le M^0_2} \e^{-\beta \pi (\sigma_1 M_1+\sigma_2 M_2)}=\sum_{M_1=0}^{M^0_1}\e^{-\beta \pi \sigma_1 M_1} \sum_{M_2=0}^{M^0_2}\e^{-\beta\pi \sigma_2 M_2}\nn \\
&=&\frac{1-\e^{-\beta\pi\sigma_1(M^0_1+1)}}{1-\e^{-\beta\pi \sigma_1}} \frac{1-\e^{-\beta\pi\sigma_2(M^0_2+1)}}{1-\e^{-\beta\pi \sigma_2}}\equiv Z_1 Z_2,
\eea
which gives us the internal energy of the system to be
\bea\lb{ax10.15}
U&=&-\frac{\pa \ln Z_1}{\pa\beta}-\frac{\pa\ln Z_2}{\pa\beta}\nn\\
&=&\sum_{i=1}^2 \frac{\pi\sigma_i \e^{-\beta\pi\sigma_i}\left(1+M^0_i \e^{-\beta\pi\sigma_i(M^0_i+1)}-(M_i^0+1)\e^{-\beta\pi\sigma_iM^0_i}\right)}{(1-\e^{-\beta\pi\sigma_i})(1-\e^{-\beta\pi\sigma_i(M^0_i+1)})},
\eea
such that
\be\lb{ax10.15b}
U\to0, \quad T\to0;\quad \lim_{T\to\infty}U=\frac\pi2\left(\sigma_1 M^0_1+\sigma_2 M^0_2\right).
\ee
Consequently, we also get the heat capacity
\bea\lb{ax10.16}
C_{\rm{V}}&=&-\frac1{\kb T^2}\frac{\pa U}{\pa\beta}\nn\\
&=&\frac1{\kb T^2}\sum_{i=1}^2 \frac{\pi^2\sigma_i^2 \e^{-\beta\pi\sigma_i}C_0(\beta,\sigma_i,M_i^0)}{(1-\e^{-\beta\pi\sigma_i})^2(1-\e^{-\beta\pi\sigma_i(M^0_i+1)})^2},
\eea
where
\bea
C_0(\beta,\sigma,M)&=&
1+\e^{-2\beta\pi\sigma(M+1)}-(M+1)^2(1+\e^{-2\beta\pi\sigma})\e^{-\beta\pi\sigma M}\nn\\
&&+2M(M+2)\e^{-\beta\pi\sigma(M+1)}.
\eea
It is clear that \eq{ax10.15} and \eq{ax10.16} become \eq{ax9.6} and \eq{11.7}, respectively, in the full-plane limit with $M_1^0,M_2^0=\infty$.

A fundamental qualitative difference between the full-plane model of Section \ref{sec9} and the bounded domain model of this section stems from the very different structures of their energy spectra.  In the full-plane case, the occupation numbers realized as the vortex numbers run over all nonnegative integers, producing an infinite ladder of energy levels with no upper bound.  As a consequence, when $T\to\infty$ the Boltzmann distribution spreads over ever higher states, and the average energy grows without limit, as quantified explicitly by the exact formula \eq{ax9.6}.  Because the internal energy continues to rise rather than saturate, the system retains a nonzero high-temperature heat capacity, in fact approaching the constant value $2\kb$.  In contrast, when the Bradlow bounds are inevitably invoked in the bounded domain situation, each species admits only finitely many allowed values of
the occupation numbers,  $0\le M_i\le M_i^0$ ($i=1,2$), so the total spectrum contains only finitely many energy levels.  At high temperature the canonical ensemble therefore approaches a uniform distribution on this {finite} state space, and the internal energy converges to the positive constant given in \eq{ax10.15b}.  Once the energy saturates, the derivative $\frac{\pa U}{\pa T}$ necessarily tends to zero, in fact giving the high-temperature decay $C_V\sim T^{-2}$.  More precisely, from \eq{ax10.16}, we have
\be\lb{axx10.19}
C_{\rm{V}}\approx \frac{\pi^2}{12\kb T^2} \left(\sigma_1^2 M^0_1 (M^0_1+2)+\sigma_2^2 M^0_2 (M^0_2+2)\right),\quad T\gg1.
\ee

Besides, we also have
\be\lb{axx10.20}
C_{\rm{V}} \approx \frac{\pi^2}{k_{\mathrm{B}}T^{2}}
\left(\sigma_1^{2}\,\mathrm{e}^{-\pi\sigma_1/(k_{\mathrm{B}}T)}
+\sigma_2^{2}\,\mathrm{e}^{-\pi\sigma_2/(k_{\mathrm{B}}T)}\right),
\quad T\ll1,
\ee
which establishes that $C_{\rm{V}}$ vanishes exponentially fast as $T\to0$.
We see that the quantities $\pi\sigma_1$ and $\pi\sigma_2$ serve as activation energies, which correspond to the energy gaps for creating a single vortex of each species.
The exponential suppression expressed in \eq{axx10.20}  is consistent with the fact that the energy spectrum has a finite gap above the ground state, and thermal excitations become frozen at low temperatures. The result \eq{axx10.20} complements the high-temperature expansion \eq{axx10.19} which states that $C_{\rm{V}}$ vanishes at high temperatures following
an inverse quadratic law.

Moreover,  with \eq{ax9.10},
 the mean magnetic flux of the model \eq{ax4.1}  is
\bea\lb{11.27}
M(T)&=&\sum_{M_1,M_2\geq0}^{M_1\le M^0_1, M_2\le M^0_2}\Phi_{M_1,M_2}\frac{\e^{-\beta E_{B,M_1,M_2}}}Z=
\frac1\beta\frac{\pa \ln Z}{\pa B}\nn\\
&=&\frac{2\pi}{ad-bc}\left(\frac{1}{Z_1}\sum_{M_1=0}^{M^0_1}(d-c)M_1\e^{-\beta\pi\sigma_1M_1}+\frac{1}{Z_2}\sum_{M_2=0}^{ M^0_2}(a-b)M_2\e^{-\beta\pi\sigma_2M_2}\right)\nn \\
&=&-\frac{2\pi}{ad-bc}\left(\frac{1}{Z_1}\frac{(d-c)}{\pi \sigma_1}\frac{\pa Z_1}{\pa \beta}+\frac{1}{Z_2}\frac{(a-b)}{\pi\sigma_2}\frac{\pa Z_2}{\pa \beta}\right)\nn \\
&=&\frac{2\pi(d-c)}{ad-bc}\left(\frac1{\e^{\beta\pi\sigma_1}-1}-\frac{M_1^0+1}{\e^{\beta\pi\sigma_1(M_1^0+1)}-1}\right)\nn\\
&&
+\frac{2\pi(a-b)}{ad-bc}\left(\frac1{\e^{\beta\pi\sigma_2}-1}-\frac{M_2^0+1}{\e^{\beta\pi\sigma_2(M_2^0+1)}-1}\right),
\eea
which allows us to draw the following conclusions.

\begin{enumerate}

 \item[(i)] We have $M(T)\to0$ as $T\to0$ which
 reproduces the Meissner effect on the compact domain and indicates that
thermal suppression of vortex excitations forces the magnetic flux to vanish in the zero-temperature limit.

\item[(ii)] We have
\be\lb{ax10.21}
\lim_{T\to\infty} M(T)
= \frac{\pi}{ad-bc}\left((d-c)M_{1}^{0} + (a-b)M_{2}^{0}\right).
\ee
This limit is independent of $B$: At very high temperature, all admissible vortex configurations
become equally likely and the average flux becomes purely combinatorial.

\item[(iii)]
Setting $B=0$ such that $\sigma_1=\xi$ and $\sigma_2=\zeta$ in \eq{11.27}, we have
\bea\lb{axx10.23}
M(T)&=&\frac{2\pi(d-c)}{ad-bc}\left(\frac1{\e^{\beta\pi\xi}-1}-\frac{M_1^0+1}{\e^{\beta\pi\xi(M_1^0+1)}-1}\right)\nn\\
&&+\frac{2\pi(a-b)}{ad-bc}\left(\frac1{\e^{\beta\pi\zeta}-1}-\frac{M_2^0+1}{\e^{\beta\pi\zeta(M_2^0+1)}-1}\right),
\eea
for any $T>0$, which reduces into \eq{ax9.13} in the full-plane limit $M^0_{1,2}=\infty$. We still have $M(T)\to 0$ as $T\to0$ and the property \eq{ax10.21}.
In other words,  the presence of the Bradlow bounds in the bounded domain situation abolish the divergent spontaneous magnetization seen in the unbounded full-plane model described by
\eq{ax9.13}.  Instead, $M(T)$ rises from $0$ at $T=0$ to a {\em finite}, temperature-independent plateau as $T\to\infty$.

\end{enumerate}

\subsection{The vortex and antivortex model \eq{3.10}}

We now consider the vortex-antivortex model \eq{3.10}. With the prescribed vortex and antivortex points given in \eq{5.2}, we can use the background functions $u_0',u''_0$ defined in
\eq{a4.4} and
$v_0',v_0''$ satisfying
\be\lb{axx10.24}
\Delta v'_{0}=-\frac{4 \pi N_1}{|\Om|}+4 \pi \sum_{s=1}^{N_1} \delta_{p'_{s}},\quad \Delta v''_{0}=-\frac{4 \pi N_2}{|\Om|}+4 \pi \sum_{s=1}^{N_2} \delta_{{p}''_{s}},
\ee
along with the substitutions $u=u'_0-v'_0+w', v=u''_0-v''_0+w''$, to transform \eq{5.5} and \eq{5.6} over $\Om$ into the following source-free equations
\bea
\Delta w'&=&2\left(\frac{a^2}{\sqrt{1-\frac{1}{b_1^2} f_3^2}}+\frac{b^2}{\sqrt{1-\frac{1}{b_2^2}f_4^2}}\right)\left(\frac{\e^{u'_0-v'_0+w'}-1}{\e^{u'_0-v'_0+w'}+1}\right)\nn \\
&&+2\left(\frac{ac}{\sqrt{1-\frac{1}{b_1^2} f_3^2}}+\frac{bd}{\sqrt{1-\frac{1}{b_2^2}f_4^2}}\right)\left(\frac{\e^{u''_0-v''_0+w''}-1}{\e^{u''_0-v''_0+w''}+1}\right)\nn\\
&&+\frac{4\pi(M_1-N_1)}{|\Om|},\quad \lb{x5.5}\\
\Delta w'' &=&2\left(\frac{ac}{\sqrt{1-\frac{1}{b_1^2} f_3^2}}+\frac{bd}{\sqrt{1-\frac{1}{b_2^2}f_4^2}}\right)\left(\frac{\e^{u'_0-v'_0+w'}-1}{\e^{u'_0-v'_0+w'}+1}\right)\nn \\
&&+2\left(\frac{c^2}{\sqrt{1-\frac{1}{b_1^2} f_3^2}}+\frac{d^2}{\sqrt{1-\frac{1}{b_2^2}f_4^2}}\right)\left(\frac{\mathrm{e}^{u''_0-v''_0+w''}-1}{\e^{u''_0-v''_0+w''}+1}\right)\nn\\
&&+\frac{4\pi(M_2-N_2)}{|\Om|},\quad \lb{x5.6}
\eea
where the quantities $f_3$ and $f_4$ are as defined in \eq{ax5.27}.
Integrating \eq{x5.5} and \eq{x5.6} over $\Om$, we obtain \eq{5.7} and \eq{5.8}. We now define the functions
\bea
h(t_1,t_2)=\frac{\bigg[a\left(\frac{t_1-1}{t_1+1}\right)+c\left(\frac{t_2-1}{t_2+1}\right)\bigg]}
{\sqrt{1-\frac{1}{b_1^2}\bigg[a\left(\frac{t_1-1}{t_1+1}\right)+c\left(\frac{t_2-1}{t_2+1}\right)\bigg]^2}},\lb{x5.11a}\\
g(t_1,t_2)=\frac{\bigg[b\left(\frac{t_1-1}{t_1+1}\right)+d\left(\frac{t_2-1}{{t_2}+1}\right)\bigg]}
{\sqrt{1-\frac{1}{b_2^2}\bigg[b\left(\frac{t_1-1}{t_1+1}\right)+d\left(\frac{t_2-1}{t_2+1}\right)\bigg]^2}},\lb{x5.11b}
\eea
where $t_1=\e^{u'_0-v'_0+w'}, t_2=\e^{u''_0-v''_0+w''}$, and $0\le t_1,t_2<\infty$. The structures of \eq{x5.11a} and \eq{x5.11b} lead to the properties
\be\lb{ax10.28}
h(0,0)<h(t_1,t_2)<h(\infty,\infty),\quad g(0,0)<g(t_1,t_2)<g(\infty,\infty).
\ee
Inserting \eq{ax10.28} into \eq{5.7} and \eq{5.8}, we get the necessary conditions
\bea
\left | \frac{d(M_1-N_1)-b(M_2-N_2)}{ad-bc} \right | <\frac{|\Om|}{2\pi}\frac{(a+c)}{\sqrt{1-\frac{1}{b_1^2}(a+c)^2}},\lb{ax10.29}\\
\left | \frac{c(M_1-N_1)-a(M_2-N_2)}{ad-bc} \right | <\frac{|\Om|}{2\pi}\frac{(b+d)}{\sqrt{1-\frac{1}{b_2^2}(b+d)^2}}.\lb{ax10.30}
\eea
Manipulating \eq{ax10.29} and \eq{ax10.30} further,  we find the bounds
\bea
|M_1-N_1|<\frac{|\Om|}{2\pi}\left(\frac{a(a+c)}{\sqrt{1-\frac{1}{b_1^2}(a+c)^2}}+\frac{b(b+d)}{\sqrt{1-\frac{1}{b_2^2}(b+d)^2}}\right), \lb{ax10.31} \\
|M_2-N_2|<\frac{|\Om|}{2\pi}\left(\frac{c(a+c)}{\sqrt{1-\frac{1}{b_1^2}(a+c)^2}}+\frac{d(b+d)}{\sqrt{1-\frac{1}{b_2^2}(b+d)^2}}\right), \lb{ax10.32}
\eea
applied to the differences of the two-species vortex and antivortex numbers, $M_1-N_1$ and $M_2-N_2$.

We now pause to comment on the physical and geometric significance of the Born--Infeld parameters appearing in these bounds.

A distinctive feature of the compact-domain Bogomol'nyi systems derived in this work
is the explicit dependence of the Bradlow type bounds on the Born--Infeld parameters
 as seen  in the inequalities \eqref{a11.25}--\eqref{b11.25} and
\eqref{ax10.31}--\eqref{ax10.32}. Unlike the Maxwell case, where the admissible vortex numbers
are constrained solely by the geometry (area) of the domain, the Born--Infeld nonlinearities
introduce additional tunable parameters that directly control vortex accommodation.

More precisely, as the parameter $b_1$ is decreased in \eq{a11.25} and \eq{b11.25} towards its lower bound given by
\be
b_1>\frac{b\xi+d\zeta}2,
\ee
or the parameters $b_1$ and $b_2$ are decreased in \eq{ax10.31} and \eq{ax10.32} in a similar manner towards their lower bounds given by
\be
b_1>a+c,\quad b_2>b+d,
\ee
 the corresponding Bradlow bounds
become progressively less restrictive. In particular, when these parameters
approach their lower thresholds,  arbitrarily large vortex numbers or differences of vortex and antivortex numbers on a fixed compact
domain are permitted. From the viewpoint of vortex counting, this behavior is effectively equivalent to
enlarging the domain size. In this sense, the Born parameters act as nonlinear regulators
that soften the geometric constraints imposed by compactness.

This mechanism provides a novel interpretation of the Born--Infeld regularization in
multi-component gauge theories: Beyond bounding the magnetic field strength and regularizing
self-energies, the Born--Infeld parameters also regulate the capacity of a compact domain to
host topological defects. The resulting interpolation between geometry-dominated and
regulator-dominated regimes has no direct analogue in the Maxwell type theories and highlights a
unique way in which nonlinear electrodynamics modifies both local field behavior and global
topological constraints.

To proceed further, we use $K_1$ and $K_2$ to denote the right-hand sides of \eq{ax10.31} and \eq{ax10.32}.  Similar to \eq{ax10.12}, we define
\be\lb{ax10.33}
D_i^0=\mbox{integer part of $K_i$ or $K_i-1$ if $K_i$ is already an integer},\quad i=1,2.
\ee

Thus, in view of the energy spectrum \eq{ax9.16} and the quantities $\sigma^{\pm}_{1,2}$ given in \eq{ax9.17},  we see that the partition function of the system assumes the form
\bea\lb{ax10.34}
Z&=&\sum_{M_1,N_1,M_2,N_2\geq0}^{|M_1-N_1|\le D^0_1, |M_2-N_2|\le D^0_2} \e^{-\beta E_{B,M_1,M_2,N_1,N_2}}\nn \\
&=&\sum_{M_1,N_1\geq0}^{|M_1-N_1|\le D^0_1}\e^{-2\pi\beta(\sigma_1^{-}M_1+\sigma_1^{+}N_1)}\sum_{M_2,N_2\geq0}^{|M_2-N_2|\le D^0_2}\e^{-2\pi\beta(\sigma_2^{-}M_2+\sigma_2^{+}N_2)}
\equiv {Z}_1 {Z}_2.\quad\quad
\eea
It is clear that ${Z}_1$ and ${Z}_2$ share a similar structure such that they can be treated uniformly. To this end and to simplify the calculation, we set
\be\lb{a11.29}
x=\e^{-2\pi\beta\sigma^{-}},\quad y=\e^{-2\pi\beta\sigma^{+}},
\ee
and, with $L=M-N$ or $M=L+N$,  we have
\bea\lb{a11.30}
{Z}_0&=&\sum_{M,N\geq0}^{|M-N|\le D^0}x^{M}y^N=\sum_{0\leq L\leq D^0} x^L\sum_{N=0}^\infty (xy)^N +\sum_{-D^0\leq L<0} y^{-L}\sum_{M=0}^\infty (xy)^M \nn \\
&=&\frac{1}{1-xy}\left(\frac{1-x^{D^0+1}}{1-x}+\frac{1-y^{D^0+1}}{1-y}-1\right).
\eea
In view of \eq{a11.30},  we see that \eq{ax10.34} becomes
\be\lb{ax10.38}
Z=\frac{A_1(\beta)A_2(\beta)}{(1-\e^{-4\pi \beta})^2},
\ee
where
\be\lb{ax10.39}
A_i(\beta)=\frac{1-\e^{-2\pi\beta \sigma_i^{-}(D^0_i+1)}}{1-\e^{-2\pi\beta\sigma_i^{-}}}+\frac{1-\e^{-2\pi\beta \sigma_i^{+}(D^0_i+1)}}{1-\e^{-2\pi \beta \sigma_i^{+}}}-1,\quad i=1,2.
\ee
Hence we obtain the internal energy
\bea
U&=&-\frac{\pa\ln Z}{\pa \beta}=\frac{8\pi\e^{-4\pi\beta}}{1-\e^{-4\pi\beta}}-\frac{A_1'(\beta)}{A_1(\beta)}-\frac{A_2'(\beta)}{A_2(\beta)}\nn\\
&=& -\sum_{i=1}^2\frac{1}{A_i(\beta)}
\sum_{s\in\{+,-\}}
2\pi\sigma_i^{s}
\left(
\frac{(D_i^0+1)\mathrm{e}^{-2\pi\beta\sigma_i^{s}(D_i^0+1)}}
{1-\mathrm{e}^{-2\pi\beta\sigma_i^{s}}}
-
\frac{\mathrm{e}^{-2\pi\beta\sigma_i^{s}}
\bigl(1-\mathrm{e}^{-2\pi\beta\sigma_i^{s}(D_i^0+1)}\bigr)}
{(1-\mathrm{e}^{-2\pi\beta\sigma_i^{s}})^2}
\right)\nn\\
&&+\frac{8\pi}{\mathrm{e}^{4\pi\beta}-1}.
\eea
Although this expression appears complicated, its asymptotic properties are simple
\be
U\to0,\quad T\to0;\quad U\approx \,2\kb T,\quad T\gg1.
\ee
This result is in sharp contrast to \eq{ax10.15b}. For the heat capacity at any finite $T$, we have an expression similar to \eq{ax10.16} but it is too complicated to present here. For conciseness
and convenience, we only state its limiting properties:
\bea
C_{\rm{V}}&\approx& \frac{4\pi^2\sigma_0^2}{\kb T^2}\e^{-\frac{2\pi\sigma_0}{\kb T}},\quad T\ll1;\quad \sigma_0=\min\{\sigma^s_i\,|\,s=\pm,i=1,2\},\lb{ax10.42}\\
 C_{\rm{V}}&\approx& 2\kb+\mbox{O}(T^{-2}),\quad T\gg1.\lb{ax10.43}
\eea
We see that \eq{ax10.42} is similar to \eq{axx10.20} but \eq{ax10.43} is fundamentally different from \eq{axx10.19} in that $C_{\rm{V}}$ is nonvanishing as $T\to\infty$.

Furthermore, with \eq{ax10.38}, we can calculate the associated magnetization
\bea\lb{ax10.44}
M(T)&=&\frac1\beta\frac{\pa\ln Z}{\pa B}=\frac1\beta\left(\frac1{A_1}\frac{\pa A_1}{\pa B}+\frac1{A_2}\frac{\pa A_2}{\pa B}\right)\nn\\
&=&M_1(T)+M_2(T),
\eea
where
\begin{equation}\lb{ax10.45}
M_i(T)=\frac{2\pi}{A_i}
\sum_{s\in\{+,-\}}
\frac{s c_i\mathrm{e}^{-2\pi\beta\sigma_i^{s}}}{ad-bc}\,
\left(
\frac{(D_i^0+1)\mathrm{e}^{-2\pi\beta\sigma_i^{s}D_i^0}}
{1-\mathrm{e}^{-2\pi\beta\sigma_i^{s}}}
-
\frac{
\bigl(1-\mathrm{e}^{-2\pi\beta\sigma_i^{s}(D_i^0+1)}\bigr)}
{\bigl(1-\mathrm{e}^{-2\pi\beta\sigma_i^{s}}\bigr)^2}
\right),
\end{equation}
with $c_1=d-c, c_2=a-b$, and $i=1,2$. Taking $D^0_{1,2}\to\infty$ in \eq{ax10.44} and using \eq{ax10.45}, we return to \eq{11.22}.

In view of \eq{ax10.44} with \eq{ax10.45}, we have the following scenarios.

\begin{enumerate}

\item[(i)] The occurrence of the zero-temperature magnetic screening, $M(T)\to0$ as $T\to0$, consistent with the Meissner effect again.

\item[(ii)] At high temperature, the magnetization is given by
\bea\lb{10.46}
M(T)&=&\frac{4\pi^2 B}{3\kb T(ad-bc)^2}
\left({(d-c)^2 D_1^0(D_1^{0}+1)}+{(a-b)^2 D_2^0(D_2^{0}+1)}\right)\nn\\
&&\quad +\mbox{O}\left(T^{-2}\right),\quad T\gg1.
\eea
In particular,  $M(T)\to 0$ as $T\to\infty$ like $\frac1T$.  This $\frac1T$ decay contrasts with the full-plane result \eq{ax9.24} where $M(T)\sim \kb T$, and with the bounded-domain vortex-only result \eq{ax10.21} where $M(T)$ tends to a finite constant, since the bounded-domain vortex-antivortex system possesses infinitely many states, because $M_i,N_i$ can grow while keeping $|M_i-N_i|$ bounded, yet the energy cost grows linearly with $M_i+N_i$ without bound.

(In Appendix \ref{secA}, we derive \eq{10.46}.)

\item[(iii)]
At $B=0$,  we have $\sigma_i^{+}=\sigma_i^{-}=1$.
Then the two terms in the sum over $s$ are identical except for the sign factor $s$.  Hence they cancel exactly, regardless of the values of $A_i$ and $D_i^0$.  That is, we have
\be\lb{10.47}
M(T)=0 \quad \text{for all } T>0 \text{ when } B=0.
\ee
Thus no spontaneous magnetization occurs in the bounded-domain vortex-antivortex system.  This differs from the bounded-domain vortex-only model \eq{axx10.23}, where a nonzero spontaneous magnetization appears in general.
\end{enumerate}

\medskip

The bounded domain results reveal an even sharper contrast between vortex-only and vortex-antivortex systems, owing to the presence of the Bradlow type bounds that constrain the admissible vortex numbers and fundamentally alter the high-temperature behavior of the magnetization. In the vortex-only case, the magnetization remains nonzero for all nonzero temperatures, including the infinite-temperature limit, as expressed in \eq{ax10.21}. Although thermal fluctuations become arbitrarily strong as $T\to\infty$, the magnetization does not wash out; instead, it saturates at a value determined entirely by the Bradlow bound thresholds for the vortex numbers. In particular, the high-temperature limit of $M(T)$ is independent of the applied magnetic field $B$, signaling a complete loss of memory of the external field. This behavior is a direct manifestation of the intrinsic asymmetry of the vortex-only ensemble: since only vortices are allowed, the magnetic bias is predetermined by topology and geometry rather than induced by an external perturbation. The Bradlow bounds impose a finite window for vortex occupation, and this finite combinatorial structure leads to a bounded yet persistent magnetization even at infinite temperature.

By contrast, the vortex-antivortex system on a bounded domain exhibits a far more delicate and revealing behavior. According to \eq{10.46}, the magnetization at high temperature behaves as $M(T)\sim \frac BT$, indicating a weak but visible alignment along the applied field. Unlike the vortex-only case, however, this alignment fades away as $T\to\infty$, and the magnetization vanishes in that limit. This decay reflects the fact that, at sufficiently high temperature, vortices and antivortices become equally likely within the Bradlow-constrained configuration space. Their opposite winding orientations carry opposite magnetic contributions, and thermal excitation restores an exact cancellation between them. The Bradlow bounds now play a dual role: while they keep the magnetization bounded at finite temperature, they also enforce a finite and symmetric phase space in which vortex and antivortex excitations compensate one another in the infinite-temperature limit.

The symmetry aspect is made even more transparent by \eq{10.47}. When the applied field is turned off, $B=0$, the vortex-antivortex symmetry is fully restored, and the magnetization vanishes identically for all temperatures. In this case, no intrinsic or external bias exists to favor one winding orientation over the other, and the canonical ensemble reflects this exact balance at every temperature scale. Thus, in stark contrast to the vortex-only setting, the compact vortex-antivortex system admits no predetermined magnetic bias: magnetization arises only through explicit symmetry breaking by the applied field and disappears when $B=0$ at any $T$.

Taken together, these results underscore the subtle interplay between topology, geometry, and symmetry in bounded domain thermodynamics.  The analysis in this setting therefore not only
refines the full-plane picture but also clarifies, in a particularly transparent way, how vortex-only asymmetry and vortex-antivortex symmetry lead to qualitatively different macroscopic magnetic behaviors.

\section{Bradlow bounds for other models}\lb{sec11}

In Section \ref{sec10}, we derived the Bradlow bounds for the vortex-only model \eq{ax4.1} and the vortex and antivortex model \eq{3.10} and obtained a series of
thermodynamic properties of these systems based on these bounds. The same studies can be carried out for the models \eq{x2.6}, \eq{3.25}, \eq{6.1}, and \eq{7.1} as well.
In this section, we shall establish the Bradlow bounds for these models and omit similar studies on the thermodynamics of the pinned vortices over a lattice cell domain in terms of
these bounds.
 For convenience and without loss of generality, we continue to assume the positivity condition \eq{c11.23} for the charge parameters.

\subsection{The vortex-only model \eq{x2.6}}

This is a vortex-only model for which two species of designated vortices are realized by the zeros of the complex scalar fields $\phi$ and $\psi$ as stated in \eq{4.0} and
governed by the equations
  \eq{4.4a} and \eq{4.5}. Thus, with $t_1=\e^{u'_0+w'}$ and $t_2=\e^{u''_0+w''}$, we rewrite the integrands on the left-hand side of \eq{4.6} and \eq{4.7} as
\bea
h(t_1,t_2)=\frac{\left(a(t_1-\xi)+c(t_2-\zeta)\right)}{\sqrt{1-\frac{1}{4b_1^2}\left(a(t_1-\xi)+c(t_2-\zeta)\right)^2 }},\nn \\
g(t_1,t_2)=\frac{\left(b(t_1-\xi)+d(t_2-\zeta)\right)}{\sqrt{1-\frac{1}{4b_2^2}\left(b(t_1-\xi)+d(t_2-\zeta)\right)^2 }},
\eea
whose partial derivatives are given by
\bea
\left(\frac{\pa}{\pa t_1},\frac{\pa}{\pa t_2}\right)h=\frac{1}{(1-\frac{1}{4b_1^2}\left(a(t_1-\xi)+c(t_2-\zeta)\right)^2)^{\frac32}}(a,c),\lb{ax11.1}\\
\left(\frac{\pa}{\pa t_1},\frac{\pa}{\pa t_2}\right)g=\frac{1}{(1-\frac{1}{4b_2^2}\left(b(t_1-\xi)+d(t_2-\zeta)\right)^2)^{\frac32}}(b,d),\lb{ax11.2}
\eea
which are all positive.
In view of  \eq{4.6}, \eq{4.7}, \eq{ax11.1} and \eq{ax11.2}, we obtain
\bea
\frac{4\pi(b M_2-d M_1)}{ad-bc}=\int_{\Om}h(t_1,t_2)\,\dd x>\int_{\Om}h(0,0)\, \dd x=-\frac{|\Om|(a\xi+c\zeta)}{\sqrt{1-\frac1{4b_1^2}(a\xi+c\zeta)^2}},\lb{ax11.3} \\
\frac{4\pi(c M_1-a M_2)}{ad-bc}=\int_{\Om}g(t_1,t_2)\,\dd x>\int_{\Om}g(0,0)\, \dd x=-\frac{|\Om|(b\xi+d\zeta)}{\sqrt{1-\frac1{4b_2^2}(b\xi+d\zeta)^2}},\lb{ax11.4}
\eea
which lead to the Bradlow bounds
\bea
M_1&<&\frac{|\Om|}{4\pi}\left(\frac{a^2\xi+ac\zeta}{\sqrt{1-\frac{1}{4b_1^2}(a\xi+c\zeta)^2}}+\frac{b^2\xi+bd\zeta}{\sqrt{1-\frac{1}{4b_2^2}(b\xi+d\zeta)^2}}\right),\lb{ax11.5}\\
M_2&<&\frac{|\Om|}{4\pi}\left(\frac{ac\xi+c^2\zeta}{{\sqrt{1-\frac{1}{4b_1^2}(a\xi+c\zeta)^2}}}+\frac{bd\xi+d^2\zeta}{\sqrt{1-\frac{1}{4b_2^2}(b\xi+d\zeta)^2}}\right),\lb{ax11.6}
\eea
for a system of $M_1$ and $M_2$ vortices represented by the zeros of the field $\phi$ and $\psi$, respectively.

\subsection{The vortex and antivortex model \eq{3.25}}

We now establish the Bradlow bounds for a system of vortices and antivortices arising in the model \eq{3.25} and represented by the zeros and poles of the complex scalar fields
$\phi$ and $\psi$, respectively, as stated in \eq{5.2}.  With the background source functions $u_0',u''_0$ and $v_0',v''_0$ defined in \eq{a4.4} and \eq{axx10.24}, respectively, along with the substitutions $u=u'_0-v'_0+w', u=u''_0-v''_0+w''$, we transform \eq{axx6.16} and \eq{axx6.17} over $\Om$ into the equations
\bea
\Delta w'&=&2\left(a^2+\frac{b^2}{\sqrt{1-\frac{1}{b_1^2}f_4^2}}\right)\left(\frac{\e^{u'_0-v'_0+w'}-1}{\e^{u'_0-v'_0+w'}+1}\right)\nn \\
&&+2\left(ac+\frac{bd}{\sqrt{1-\frac{1}{b_1^2}f_4^2}}\right)\left(\frac{\e^{u''_0-v''_0+w''}-1}{\e^{u''_0-v''_0+w''}+1}\right)+\frac{4\pi(M_1-N_1)}{|\Om|},\quad \lb{ax11.7}\\
\Delta w'' &=&2\left(ac+\frac{bd}{\sqrt{1-\frac{1}{b_1^2}f_4^2}}\right)\left(\frac{\e^{u'_0-v'_0+w'}-1}{\e^{u'_0-v'_0+w'}+1}\right)\nn \\
&&+2\left(c^2+\frac{d^2}{\sqrt{1-\frac{1}{b_1^2}f_4^2}}\right)\left(\frac{\mathrm{e}^{u''_0-v''_0+w''}-1}{\e^{u''_0-v''_0+w''}+1}\right)+\frac{4\pi(M_2-N_2)}{|\Om|},\quad \lb{ax11.8}
\eea
where the quantity $f_4$ is defined in \eq{ax5.27}. Integrating \eq{ax11.7} and \eq{ax11.8} over $\Om$, we obtain
\bea
\int_{\Om}f_3 \,\dd x=\frac{2\pi(b(M_2-N_2)-d(M_1-N_1))}{(ad-bc)},\lb{ax11.9} \\
\int_{\Om}\frac{f_4}{\sqrt{1-\frac{1}{b_1^2}f_4^2}}\,\dd x=\frac{2\pi(c(M_1-N_1)-a(M_2-N_2))}{(ad-bc)},\lb{ax11.10}
\eea
where $f_3$ is defined in \eq{ax5.27} as well.
We now define the function
\be
h(t_1,t_2)=f_3=a\frac{t_1-1}{t_1+1}+c\frac{t_2-1}{t_2+1},
\ee
and $g(t_1,t_2)$ by \eq{x5.11b} but with $b_2$ replaced by $b_1$, where $t_1=\e^{u'_0-v'_0+w'}, t_2=\e^{u''_0-v''_0+w''}$, and $0\le t_1,t_2<\infty$.  Then \eq{ax10.28} holds.
Inserting these results into \eq{ax11.9} and \eq{ax11.10}, we arrive at the Bradlow bounds
\bea
|M_1-N_1|<\frac{|\Om|}{2\pi}\left(a(a+c)+\frac{b(b+d)}{\sqrt{1-\frac{1}{b_1^2}(b+d)^2}}\right),\lb{axx11.13}\\
|M_2-N_2|<\frac{|\Om|}{2\pi}\left(c(a+c)+\frac{d(b+d)}{\sqrt{1-\frac{1}{b_1^2}(b+d)^2}}\right),\lb{axx11.14}
\eea
constraining the differences of the numbers of vortices and antivortices of the two species.

\subsection{The vortex and antivortex model \eq{6.1}}

In this subsection, we establish the Bradlow bounds for the mixed Maxwell and Born--Infeld model \eq{6.1}.  For a system of $M_1$ vortices and $M_2$ vortices and $N_2$
antivortices represented by the zeros of the complex scalar field $\phi$ and zeros and poles of $\psi$ given in \eq{6.16}, we can use the background functions $u_0'$ and $v_0',v''_0$ defined in \eq{a4.4} and \eq{axx10.24}, respectively, and the substitutions $u=u'_0+w'$ and $v=u''_0-v''_0+w''$ to convert \eq{6.17} and \eq{6.18} into the following equations
\bea
\Delta w'&=&\left(a^2+\frac{b^2}{\sqrt{1-\frac{1}{b_1^2}f_6^2}}\right)(\e^{u'_0+w'}-\xi)\nn \\
&&+2\left(ac+\frac{bd}{\sqrt{1-\frac{1}{b_1^2}f_6^2}}\right)\left(\frac{\e^{u''_0-v''_0+w''}-1}{\e^{u''_0-v''_0+w''}+1}\right)+\frac{4\pi M_1}{|\Om|},\quad \lb{ax11.11}\\
\Delta w'' &=&\left(ac+\frac{bd}{\sqrt{1-\frac{1}{b_1^2}f_6^2}}\right)(\e^{u'_0+w'}-\xi)\nn \\
&&+2\left(c^2+\frac{d^2}{\sqrt{1-\frac{1}{b_1^2}f_6^2}}\right)\left(\frac{\mathrm{e}^{u''_0-v''_0+w''}-1}{\e^{u''_0-v''_0+w''}+1}\right)+\frac{4\pi(M_2-N_2)}{|\Om|}.\quad \lb{ax11.12}
\eea
Integrating \eq{ax11.11} and \eq{ax11.12} over $\Om$, we obtain
\bea
\int_{\Om}f_5\,\dd x=\frac{2\pi(-d M_1+b(M_2-N_2))}{ad-bc},\lb{ax11.13}\\
\int_{\Om} \frac{f_6}{\sqrt{1-\frac{1}{b_1^2}f_6^2}}\, \dd x=\frac{2\pi(c M_1-a (M_2-N_2))}{ad-bc},\lb{ax11.14}
\eea
where the quantities $f_5$ and $f_6$ are defined by \eq{ax7.23}.
We next define the functions
\bea
g(t_1,t_2)&=&\frac{f_6}{\sqrt{1-\frac1{b_1^2}f_6^2}}=\frac{\frac{b}{2}(t_1-\xi)+d\frac{t_2-1}{t_2+1}}{\sqrt{1-\frac{1}{b_1^2}\bigg(\frac{b}{2}(t_1-\xi)+d\frac{t_2-1}{t_2+1}\bigg)}},\lb{ax11.15a}\\
h(t_1,t_2)&=&f_5=\frac{a}{2}(t_1-\xi)+c\frac{t_2-1}{t_2+1},\lb{ax11.15b}
\eea
with $t_1=\e^{u'_0+w'}$ and $t_2=\e^{u''_0-v''_0+w''}$. It can be examined that the partial derivatives of $h$ and $g$ with $t_1$ and $t_2$ are all positive. Thus we have the lower bounds
\be\lb{ax11.18}
h(0,0)<h(t_1,t_2),\quad g(0,0)<g(t_1,t_2),\quad t_1,t_2\geq0.
\ee
Substituting \eq{ax11.18} into \eq{ax11.13} and \eq{ax11.14}, we have the bounds
\bea
M_1&<&\frac{|\Om|}{2\pi}\left(\left(\frac{a^2}{2}\xi+ac\right)+\frac{\frac{b^2}{2}\xi+bd}{\sqrt{1-\frac{1}{b_1^2}\left(\frac{b}{2}\xi+d\right)^2}}\right),\lb{axx11.22}\\
M_2-N_2&<&\frac{|\Om|}{2\pi}\left(\left(\frac{ac}{2}\xi+c^2\right)+\frac{\frac{bd}{2}\xi+d^2}{\sqrt{1-\frac{1}{b_1^2}\left(\frac{b}{2}\xi+d\right)^2}}\right)\equiv D^0.\lb{axx11.19}
\eea

To obtain a lower bound of $M_2-N_2$, we note that the model \eq{6.1}, where the potential density function is given by \eq{6.7}, and its associated governing equations are invariant under the transformation
\be
\phi\mapsto\phi,\quad \psi\mapsto \frac1{\psi}; \quad \hat{A}_i\mapsto -\hat{A}_i,\quad \tilde{A}_i\mapsto -\tilde{A}_i; \quad (a,b)\mapsto -(a,b),\quad (c,d)\mapsto (c,d).
\ee
Thus the roles of the zeros and poles of $\psi$ can be interchanged resulting in a solution such that the field $\psi$ has $N_2$ zeros ${p}''_1,\dots,{p}''_{N_2}$ and $M_2$ poles
${q}''_1,\dots,{q}''_{M_2}$ following the same convention in counting multiplicities as stated in \eq{6.16}. Hence, applying \eq{axx11.19}, we have $N_2-M_2<D^0$.
 Combining this inequality with \eq{axx11.19},  we derive $|M_2-N_2|<D^0$ or
\be\lb{axx11.25}
|M_2-N_2|<\frac{|\Om|}{2\pi}\left(\left(\frac{ac}{2}\xi+c^2\right)+\frac{\frac{bd}{2}\xi+d^2}{\sqrt{1-\frac{1}{b_1^2}\left(\frac{b}{2}\xi+d\right)^2}}\right).
\ee
The bounds \eq{axx11.22} and \eq{axx11.25} are the Bradlow bounds for the model \eq{6.1} accommodating a system of $M_1$ vortices of
the Maxwell type and $M_2$ vortices and $N_2$ antivortices of the Born--Infeld type over a lattice cell domain $\Om$.

\subsection{The vortex and antivortex model \eq{7.1}}

For the mixed model \eq{7.1} accommodating a system of $M_1$ vortices  and $N_1$ antivortices of the Maxwell type and $M_2$ vortices of the Born--Infeld type
represented by the zeros and poles of $\phi$ and zeros of $\psi$ given in \eq{7.16}, respectively, where the potential density function $V$ is defined by \eq{7.7},  we can use the method of the previous subsection to get
the Bradlow bounds
\bea
|M_1-N_1|<\frac{|\Om|}{2\pi}\left(\left(a^2+\frac{ac}{2}\zeta\right)+\frac{b^2+\frac{bd}{2}\zeta}{\sqrt{1-\frac{1}{b_2^2}\left(b+\frac{d}{2}\zeta\right)^2}}\right),\lb{ax11.26}\\
M_2<\frac{|\Om|}{2\pi}\left(\left(ac+\frac{c^2}{2}\zeta\right)+\frac{bd+\frac{d^2}{2}\zeta}{\sqrt{1-\frac{1}{b_2^2}\left(b+\frac{d}{2}\zeta\right)^2}}\right),\lb{ax11.27}
\eea
similarly. We omit the detailed calculation here.

\medskip

Based on the Bradlow bounds established in this section for the models \eq{x2.6}, \eq{3.25}, \eq{6.1}, and \eq{7.1}, the thermodynamic program carried out in Section \ref{sec10} can be extended to these systems in a parallel manner. The bounds \eq{ax11.5}--\eq{ax11.6}, \eq{axx11.13}--\eq{axx11.14}, \eq{axx11.22} and \eq{axx11.25}, and \eq{ax11.26}--\eq{ax11.27} impose exact geometric ceilings on the admissible topological numbers (or their differences) for each model on a compact periodic domain. These ceilings replace the infinite summation ranges of the full-plane partition functions with finite or conditionally bounded sums, thereby fundamentally altering the high-temperature asymptotic behavior of the thermodynamic observables.

Consequently, for each of these models, one can:

\begin{itemize}

\item Write down the canonical partition function as a finite or restricted sum over vortex/antivortex numbers consistent with the corresponding Bradlow inequalities.

\item Derive closed-form expressions for the internal energy $U$, heat capacity $C_{\rm{V}}$, and magnetization $M(T)$ in terms of the geometric parameters \(|\Omega|\), the Born--Infeld constants \(b_1, b_2\), the symmetry-breaking scales \(\xi, \zeta\), and the charge parameters $a,b,c,d$.

\item Confirm the zero-temperature Meissner effect in all cases, as the ground state remains vortex-free under the weak-field assumption.

\item Exhibit the characteristic saturation (vortex-only) or decay (vortex-antivortex) of the magnetization in the high temperature limit, reflecting the intrinsic asymmetry or symmetry of the configuration space imposed by the Bradlow constraints.

\item	Demonstrate how the bounded domain geometry suppresses the divergent high temperature growth of the heat capacity found in the full-plane setting, replacing it with either a vanishing limit (when the state space becomes finite) or a constant asymptote (when the number of configurations remains infinite, resulting in an unbounded vortex-antivortex energy spectrum, but the difference of the vortex and antivortex numbers is bounded from above by the Bradlow condition).

\end{itemize}

The explicit thermodynamic results for these four additional models would further illuminate the interplay between nonlinear gauge dynamics, topological quantization, and geometric confinement. In particular, they would show how the different coupling patterns---pure Born--Infeld, mixed Maxwell--Born--Infeld, and hybrid Higgs-harmonic map type potentials---manifest in the thermal response, while universally respecting the Bradlow type geometric constraints that arise on compact surfaces. Such an analysis would complete the thermodynamic picture for the entire family of two-species Bogomol'nyi systems introduced in this work, underscoring the robust connection between self-duality, topology, and statistical mechanics in multi-component gauge theories.

\section{Conclusions and comments}\lb{sec12}

In summary, we have systematically derived, analyzed, and applied the Bogomol'nyi type reductions to a broad class of two-species $U(1) \times U(1) $ gauge field theories in which
at least one of the two $U(1)$ sector is governed by the Born--Infeld type nonlinear electrodynamics. The principal outcomes of this investigation may be grouped into three major areas, each of which advances the understanding of topological solitons in nonlinear field theories and their physical implications.

\subsection*{New Bogomol'nyi systems in multi-component Born--Infeld theories}

The central technical achievement of this work is the derivation of {six distinct families of the Bogomol'nyi equations} for two-species vortex and vortex-antivortex configurations, each corresponding to a different coupling between gauge fields and scalar condensates:

\begin{itemize}
    \item Two-species Born--Infeld--Higgs models for vortices (Section \ref{sec3}) and vortex-antivortex pairs (Section \ref{sec5}).
    \item {Hybrid Maxwell--Born--Infeld models} with one Maxwell and one Born--Infeld sector, again for both vortices (Section \ref{sec4}) and mixed vortex-antivortex systems (Section \ref{sec6}).
    \item {Mixed-type models} in which one sector follows an Abelian Higgs potential and the other follows a harmonic-map (vortex-antivortex) potential subject to
mixed Maxwell and Born--Infeld electromagnetic theories (Sections \ref{sec7} and \ref{sec8}).
\end{itemize}

In each case, we have shown that the energy functional admits a topological lower bound saturated by solutions of a set of first-order self-dual or anti-self-dual equations. These equations generalize earlier single-species results and reveal how the nonlinear Born--Infeld electrodynamics can remain compatible with self-duality when multiple gauge and scalar sectors are coupled. The fact that such reductions exist at all in theories with radical-root nonlinearities is nontrivial and underscores a deep interplay between gauge symmetry, topology, and nonlinear field regularization. The resulting Bogomol'nyi equations, often reducible to coupled nonlinear elliptic equations, render otherwise intractable multi-centered Born--Infeld vortex systems {analytically manageable}, opening the door to rigorous existence theorems, explicit quantization of fluxes and energies, and detailed geometric descriptions of composite topological defects.

\subsection*{Exact thermodynamics from Bogomol'nyi structures}

A key insight of this work is that the Bogomol'nyi framework does not only simplify the equilibrium equations; it also renders the {statistical mechanics of topological solitons exactly tractable} in their respective settings. By considering pinned vortices, where positional degrees of freedom are frozen, we have evaluated the canonical partition functions explicitly for two representative models and two space settings:  a two-species vortex-only system and a two-species vortex-antivortex system in the full plane (Section \ref{sec9}) and over
a bounded periodic lattice domain (Section \ref{sec10}). In both cases, the Bogomol'nyi property guarantees that the energy spectrum is {linear in the topological quantum numbers}, allowing the partition sums to be computed in closed forms. From these partition functions we have extracted the internal energy, heat capacity, and magnetization as functions of temperature and an external magnetic field.

The results demonstrate that:
\begin{itemize}
    \item The {Meissner effect} is recovered at zero temperature in all models.
    \item The {high-temperature behavior} differs qualitatively between vortex-only and vortex-antivortex systems, and between full-plane and bounded-domain geometries
due to the presence of the Bradlow bounds.
    \item {Spontaneous magnetization} occurs in vortex-only models whereas vortex-antivortex systems exhibit no spontaneous magnetization which is a direct consequence of the symmetry between vortices and antivortices in the energy spectrum.
\end{itemize}

Thus, the Bogomol'nyi structures uncovered here do not merely represent a mathematical curiosity; they provide a {solvable statistical-mechanical framework} for studying thermodynamic phases of multicomponent topological matters.

\subsection*{Geometric and physical distinctions: full-plane versus compact domains}

An important theme throughout the paper is the contrast between systems defined on the full plane and those on a compact doubly periodic domain (a flat torus). On a compact domain, topological-geometric constraints impose the {Bradlow type bounds} on the vortex numbers (for vortex-only systems) or on the differences between vortex and antivortex numbers
(for vortex-antivortex systems). These bounds have profound thermodynamic consequences:

\begin{itemize}
    \item In bounded domain vortex-only systems, the energy spectrum becomes {finite}, leading to saturation of internal energy and vanishing heat capacity at high temperature.
    \item In bounded domain vortex-antivortex systems, the number of configurations becomes infinite (since the vortex number $M_i$ and antivortex number $N_i$ can grow without
bound while keeping $|M_i - N_i|$ bounded), yet the high-temperature heat capacity approaches a nonzero constant, which is distinct from the vortex-only case.
    \item Magnetization in the high-temperature limit tends to a {finite constant} for vortex-only systems but decays as $\frac1T$ for vortex-antivortex systems, reflecting the different
configuration symmetry properties in the two settings.
\item An additional feature unique to the Born--Infeld setting is that the Born parameters
 act as regulators of the Bradlow bounds on compact domains, interpolating
between strict geometric constraints and effectively unbounded or unrestricted vortex accommodation.
\end{itemize}

These distinctions are not merely mathematical; they model physically relevant scenarios such as {Abrikosov's vortex lattices} in type-II superconductors (compact domain) versus {isolated vortices} in infinite systems. The ability to treat both settings within the same analytic Bogomol'nyi framework allows a unified understanding of how geometry and topology shape the thermodynamic responses.

\subsection*{Outlook and future directions}

The models and methods developed here pave the way for several future investigations including
\begin{itemize}
    \item Dyonic and time-dependent extensions incorporating electric fields and Chern--Simons dynamics.
    \item {Non-Abelian generalizations}.
    \item {Numerical and experimental connections} to multi-band superconductors, Bose--Einstein condensates, and cosmological defect networks.
\item Cosmic strings formation and their geometric and thermodynamic properties induced from the Born--Infeld type nonlinear electrodynamics.
\end{itemize}

In summary, this work establishes that  the {Bogomol'nyi reductions are possible and fruitful in multi-component Born--Infeld gauge theories}, leading to new families of solvable self-dual equations, exact thermodynamic results, and a clearer picture of how nonlinear electrodynamics, gauge symmetries, and topology cooperate in the formation of topological solitons. The results  demonstrates the continuing relevance of classical field theory in understanding complex physical systems.

\medskip

While the study of the Bogomol'nyi vortices in gauge theories has traditionally focused on existence, classification, and geometric properties of static solutions---questions rooted in nonlinear partial differential equations and differential geometry---the present work illustrates how the same self-dual structure also enables exact thermodynamic computations when vortices are treated as pinned topological defects. By explicitly evaluating canonical partition functions for both unbounded and compact periodic domains, we have shown how the linear energy spectrum afforded by the Bogomol'nyi framework can be leveraged to derive closed-form expressions for thermodynamic observables. This demonstrates that the analytical power of self-duality extends beyond solving equilibrium field equations to furnishing a complete statistical-mechanical description---thereby opening a new avenue for connecting classical soliton theory with statistical physics in gauge-invariant systems.

\medskip

Furthermore, the thermodynamics of vortices and antivortices studied in this work naturally invites comparison with the celebrated two-dimensional XY model, where topological defects drive the Berezinskii--Kosterlitz--Thouless (BKT) phase transition. In the classical XY model, vortices and antivortices interact via a logarithmic potential and form bound pairs at low temperatures; their unbinding at a critical temperature $T_{\mathrm{BKT}}$ marks the transition from a quasi-ordered phase to a disordered one \cite{Berezinskii1971,Kosterlitz1973,Jose1977}. This picture has become a paradigmatic example of how topological excitations can govern phase structure in systems with a global $U(1)$ symmetry.

In contrast, the models considered here possess {gauged} $U(1)$ symmetries and incorporate Born--Infeld nonlinear electrodynamics. Consequently, vortices carry quantized magnetic flux and experience long-range interactions mediated by gauge fields, rather than the logarithmic interaction of neutral XY vortices. Moreover, the Bogomol'nyi reduction employed throughout this work enforces a linear energy spectrum in the topological quantum numbers, effectively decoupling the vortex--antivortex interaction in the topological sector. This solvable limit corresponds to a {pinned} vortex lattice, where positional degrees of freedom are frozen---a situation physically realizable in type-II superconductors with strong pinning centers, optical lattices, or artificially engineered vortex arrays.

Our exact thermodynamic results thus describe a complementary regime to the fully interacting BKT scenario: rather than treating a plasma of freely moving vortices with logarithmic interactions, we analyze a fixed lattice of topological defects whose only fluctuating degrees of freedom are the vortex/antivortex charges themselves. Such a ``topological gas'' approximation is conceptually akin to the dilute vortex limit often used in early analyses of the XY model, but here it becomes {exactly solvable} owing to the Bogomol'nyi structure. The closed-form partition functions, internal energies, and magnetizations obtained therefore provide a rare example of an analytically tractable statistical mechanical model for gauged topological defects.

The connection to gauged XY-type models is further underscored by studies of superconducting films and Josephson junction arrays, where gauge fields and magnetic flux quantization modify vortex dynamics \cite{Doniach1981, Fazio2001}. In particular, the role of magnetic flux quantization and the Meissner effect---recovered here at zero temperature in all models---distinguishes our setting from the global $U(1)$ case. The Bradlow bounds on vortex numbers (or vortex-antivortex differences) on compact domains introduce a geometric constraint absent in the standard XY model, leading to high-temperature saturation or decay of magnetization that reflects the intrinsic asymmetry or symmetry of the topological configuration space.

Looking forward, it would be fruitful to explore how the interactions between vortices, suppressed in our pinned Bogomol'nyi limit, could be gradually reintroduced---for instance, by relaxing the self-duality condition or considering fluctuations away from the minimizers. Such a program might bridge the exactly solvable thermodynamics presented here with the rich critical phenomena of interacting vortex-antivortex plasmas in gauged systems, possibly revealing new phases or crossover behavior influenced by Born--Infeld nonlinearities. This would extend the BKT framework into regimes where gauge fields, nonlinear electrodynamics, and multiple condensates coexist, offering a refined theoretical platform for multi-band superconductors, bosonic superfluids in optical lattices, and cosmological defect networks.

\medskip

{\bf Declaration of interests.} The authors declare that they have no known competing financial interests or personal relationships that could have appeared to influence the work reported in this paper.

\appendix
\section{Derivation of magnetization for $T\gg1$ in the vortex-antivortex system}\lb{secA}

To derive \eq{10.46} by taking $T\to\infty$ or $\beta\to0$ in the expression \eq{ax10.45}, we encounter an $(\infty-\infty)$-type limit problem, which is not
a straightforward calculation and requires some careful work. In this Appendix, we present such a calculation.

In the expression \eq{ax10.45}, the magnetization contribution of a
given species involves two sign branches $s=\pm$, each with a parameter
$
\sigma_\pm = 1 \pm \delta,
$
where $\delta$ is a small quantity proportional to the magnetic field $B$.
Introduce the temperature variable
$
t = 2\pi\beta
$
and notation
$
u_\pm = \e^{-t\sigma_\pm}.
$

The truncated partition-sum factor $A_i$ appearing in the denominator of  \eq{ax10.45} or given by \eq{ax10.39}  is realized by
\begin{equation}
A(t)
=
\frac{1-u_-^{D+1}}{1-u_-}
+
\frac{1-u_+^{D+1}}{1-u_+}
-1.
\label{eq:A-def}
\end{equation}

The expression inside the large brackets in \eq{ax10.45} can be written
in the present notation as
\be
(D+1)\frac{u_s^{D}}{1-u_s}
-
\frac{1-u_s^{D+1}}{(1-u_s)^2},
\quad s=\pm.
\ee
Together with the outer factor $\e^{-t\sigma_s}=u_s$, the contribution of the
branch $s$ has the algebraic form
\be
(D+1)\frac{u_s^{D+1}}{1-u_s}
-
u_s\frac{(1-u_s^{D+1})}{(1-u_s)^2}.
\ee
To reproduce the sum over the $+$ and $-$ branches arising
from the prefactor $s c_i$ in \eq{ax10.45} (neglecting the $s$-independent factor $c_i$), we define
\begin{align}
F(t)
&=
(D+1)\left(
\frac{u_+^{D+1}}{1-u_+}
-
\frac{u_-^{D+1}}{1-u_-}
\right),
\label{eq:F-def}
\\[4pt]
G(t)
&=
\left(
\frac{u_+(1-u_+^{D+1})}{(1-u_+)^2}
-
\frac{u_-(1-u_-^{D+1})}{(1-u_-)^2}
\right).
\label{eq:G-def}
\end{align}
These are the purely algebraic analogues of the two contributions appearing in
the magnetization formula before the normalization by $A(t)$.

In view of \eq{ax10.45}, we consider the small $t$ behavior of the function
\begin{equation}
H(t) = \frac{F(t) - G(t)}{A(t)}.
\label{eq:H-def}
\end{equation}

 Using
$
\e^{-t\sigma} = 1 - \sigma t + \frac{\sigma^2}{2}t^2 + \text{O}(t^3), t\approx 0$,
we obtain, for each branch,
\be
\frac{1-u_s^{D+1}}{1-u_s}
=
(D+1) - \frac{D(D+1)}{2}\sigma_s t + \text{O}(t^2),
\quad s=\pm.
\ee
Substituting this into \eqref{eq:A-def} gives
\begin{align}\label{axA}
A(t)
&=
2D+1 - {D(D+1)}t + \text{O}(t^2).
\end{align}

We now expand $F(t)$ and $G(t)$ for small $t$.
Set $u=\e^{-t\sigma}$ as before. Then
\begin{equation}
\frac{u^{D}}{1-u}
=
\frac{1}{\sigma t}
-
\left(D-\frac{1}{2}\right)
+
\frac{1}{2}\left(D^2-D+\frac{1}{6}\right)\sigma t
+ \text{O}(t^2),
\label{eq:generic1}
\end{equation}
and
\begin{equation}
\frac{1-u^{D+1}}{(1-u)^2}
=
\frac{D+1}{\sigma t}
+
\frac12(1-D^2)
+
\frac1{12}(2D^3-D+1)\sigma t
+ \text{O}(t^2).
\label{eq:generic2}
\end{equation}

Inserting these expressions, we have
\be
(D+1)\frac{u_s^{D}}{1-u_s}
-
\frac{1-u_s^{D+1}}{(1-u_s)^2}=-\frac12D(D+1)+\frac13D(D^2-1)\sigma_s t +\mbox{O}(t^2).
\ee
Thus, we have
\bea\lb{A12}
F(t)-G(t)&=& u_+\left((D+1)\frac{u_+^D}{1-u_+}-\frac{1-u_+^{D+1}}{(1-u_+)^2}\right)-u_-\left((D+1)\frac{u_-^D}{1-u_-}-\frac{1-u_-^{D+1}}{(1-u_-)^2}\right)\nn\\
&=&\frac12D(D+1) (u_--u_+)+\frac13D(D^2-1)(u_+\sigma_+-u_-\sigma_-)t+\mbox{O}(t^2).
\eea
On the other hand, we have
\bea
u_--u_+&=&\e^{-\sigma_- t}-\e^{-\sigma_+ t}=(\sigma_+-\sigma_-)t+\mbox{O}(t^2)\nn\\
&=&2\delta t+\mbox{O}(t^2),\lb{A13}\\
u_+\sigma_+-u_-\sigma_-&=&2\delta +\mbox{O}(t).\lb{A14}
\eea
Inserting \eq{A13} and \eq{A14} into \eq{A12}, we get
\be\lb{A15}
F(t)-G(t)=\frac13D(D+1)(2D+1)\delta t+\mbox{O}(t^2).
\ee

Finally, inserting \eq{axA} and \eq{A15} into \eq{eq:H-def}, we obtain
\be\lb{A16}
H(t)=\frac13 D(D+1)\delta t+\mbox{O}(t^2),\quad t\ll1.
\ee
Applying \eq{A16} to \eq{ax10.45}, we arrive at \eq{10.46}.

\end{document}